\documentclass{jpp}
\usepackage{graphicx}
\usepackage{hyperref}

\usepackage{etoolbox}

\providecommand*{\Autoref}[1]{%
  \begingroup
  \def\chapterautorefname{Chapter}%
  \def\sectionautorefname{Section}%
  \def\subsectionautorefname{Subsection}%
  \def\subsubsectionautorefname{Subsubsection}%
  \def\figureautorefname{Figure}%
  \def\tableautorefname{Table}%
  \def\equationautorefname{Equation}%
  \def\lstnumberautorefname{Line}
  \autoref{#1}%
  \endgroup
}

\usepackage{amsmath}
\DeclareMathOperator*{\argmin}{argmin}

\usepackage{caption}
\usepackage{subcaption}
\usepackage{xcolor}
\definecolor{lightskyblue}{RGB}{135, 206, 235}
\usepackage{multirow}
\usepackage{bm}

\usepackage[utf8]{inputenc}
\usepackage[T1]{fontenc}
\usepackage{amsmath}
\usepackage{algorithm}
\usepackage{algpseudocode}

\shorttitle{Deflation Techniques for Stellarator Equilibrium and Optimization}
\shortauthor{D. Panici et al.}

\author{Dario Panici\aff{1}
  \corresp{\email{dpanici@princeton.edu}}, Byoungchan Jang\aff{2}, Rory Conlin\aff{2}
, Daniel Dudt\aff{3}, Yigit Gunsur Elmacioglu\aff{1}, Egemen Kolemen\aff{1}\corresp{\email{ekolemen@princeton.edu}}}

\affiliation{\aff{1}Princeton University, Princeton, New Jersey 08544, USA
\aff{2}Institute for Research in Electronics and Applied Physics, University of Maryland, College
Park, MD, 20742, USA
\aff{3}Thea Energy, USA}

\title{Deflation Techniques for Stellarator Equilibrium and Optimization}
\date{January 2026}

\begin{document}
\maketitle

\begin{abstract}
Stellarator optimization is a multi-objective, non-convex problem characterized by a complex objective landscape containing many local minima. The solution resulting from a single optimization is highly sensitive to factors such as the initial guess, objective weights, and the optimization method employed. However, merely varying these factors does not guarantee that a physically distinct minimum will be found; optimizations often fail to converge to good minima or simply return to the same or very similar local minima despite large-scale parameter scans. This paper presents a novel application of deflation methods to effectively explore this landscape. By modifying the objective function to penalize and "deflate" away already-found solutions, this technique encourages the optimizer towards attractive, distinct new minima while using a single initial guess and optimization setup. We provide a primer on deflation for nonlinear systems and non-convex optimization before applying it to non-axisymmetric equilibrium and stellarator optimization problems. Key results include the discovery of families of global equilibria with similar core characteristics and the convergence to helical core equilibria without prescient initial guesses. Furthermore, we demonstrate that augmenting stage-one stellarator and stage-two coil optimization with deflation constraints readily produces multiple high-quality, distinct solutions, establishing the method's efficacy and ease of use.

\end{abstract}

\section{Introduction}

Nuclear fusion energy is a promising potential carbon-free energy source. Attaining fusion on Earth through the toroidal magnetic confinement approach involves using strong magnetic fields to confine hot plasmas in toroidally-shaped vessels. In designing a magnetically-confined nuclear fusion reactor, care must be taken to ensure that the plasma is well-contained and not lost, which hinders energy production and can cause damage to the device walls. The tokamak, a leading toroidal magnetic fusion concept, attains excellent neoclassical confinement of the plasma due to its axisymmetry, but this comes at the cost of requiring significant plasma current to provide the confining rotational transform, current which is conventionally driven inductively. Stellarators, on the other hand, attain rotational transform through 3-D shaping of the externally-generated magnetic field, dropping the requirement of large plasma currents and more readily enabling stable, steady state operation \citep{boozer_physics_2005,helander_theory_2014}. However, the loss of a physical symmetry direction means particles are no longer as well-confined as they are in tokamaks, and so stellarators require careful optimization to recover good confinement \citep{boozer_transport_1983}. Beyond confinement, many other considerations must be taken into account when designing a stellarator, such as stability, coil complexity, and fast particle losses. In sum, stellarator optimization is a multi-objective problem with many degrees of freedom \citep{landreman_efficient_2016}, and typically one deals with nonconvex objective landscapes with multiple local minima.

Even in the ideal MHD equilibrium problem alone \citep{freidberg_ideal_2014}, at finite pressure it has been observed that multiple solutions to a given set of fixed-boundary inputs (last closed flux surface, pressure, and rotational transform profiles) can exist \citep{cooper_tokamak_2010, wingen_use_2018, rakha_shear_2019}, with the obtained solution depending on the initialization of the equilibrium geometry in the equilibrium solver (such a VMEC \citep{hirshman_steepestdescent_1983} or \texttt{DESC} \citep{dudt_desc_2020, panici_desc_2023, conlin_desc_2023, dudt_desc_2023}). Additionally, in recent work \citep{panici_extending_2025}, a  new approach to solving for ideal MHD equilibria consistent with near-axis-expansion (NAE) solutions has been shown to be under-constrained compared to the traditional fixed-boundary problem, which would logically mean that multiple solutions may exist in this case as well. \par
In each of these situations, the typical approaches to find multiple distinct minima include repeatedly solving while varying the initial condition (sometimes with the use of a reduced model like the NAE \citep{garren_magnetic_1991, landreman_mapping_2022,rodriguez_constructing_2023}), or solving from the same initial condition but while varying some hyperparameters in the objective function (such as objective targets or relative weights between objectives)\citep{bindel_understanding_2023,laia_data-driven_2026}. 

This paper introduces the application of deflation techniques to find multiple solutions to non-axisymmetric equilibrium and stellarator optimization problems. In \Autoref{sec:deflation-background}, a brief introduction to deflation methods will be given, as well as their applications to nonlinear system solving and nonconvex optimization. \Autoref{sec:finding-multiple-solns-equilibrium} then will apply these techniques to find multiple equilibrium solutions to the near-axis-constrained equilibrium problem \citep{panici_extending_2025} and to the helical core equilibrium problem \citep{cooper_tokamak_2010}, showing its utility in finding distinct solutions when multiple valid equilibrium states may exist. \Autoref{sec:finding-multiple-solns-optimization} utilizes deflation in stage-one stellarator optimization and stage-two coil optimization, to demonstrate that deflation can effectively find multiple attractive local minima to these nonconvex, multi-objective optimization problems, as well as introduce and motivate a novel form of multiple-deflation. Finally, \Autoref{sec:conclusions} will conclude with a discussion of the results, as well as ways to improve upon this work. While deflation methods have been applied in tokamak literature to find multiple solutions to the Grad-Shafranov equations \citep{ham_multiple_2024, pentland_multiple_2025}, to the authors' knowledge this work marks the first application of deflation to stellarators.

\section{Deflation Methods}\label{sec:deflation-background}

\subsection{Application to Nonlinear System Solving}
When dealing with nonlinear equations (or systems of nonlinear equations), multiple solutions may exist (e.g. $cos(x) = x^2$). Typically one employs a rootfinding algorithm such as Newton's method to solve such equations, which can effectively find single roots, but is limited in that it will only find a single root from a single given initial guess. Furthermore, which root is found can depend on the initial guess used for the rootfinding, \textcolor{black}{and for certain initial guesses the algorithm may fail to converge or even diverge}. In \cite{Wilkinson1963}, deflation methods were introduced to find multiple roots to polynomial equations from the same initial guess through the introduction of a barrier around known roots, which \cite{brow_deflation_1971}, extended to systems of nonlinear equations. In deflation methods, once a single solution to the system of equations is found, the system is modified in such a way that the rootfinding may be re-applied to yield additional, distinct solutions to the system. More concretely, let  

\begin{equation}
    \bm{F}(\bm{x}) = 0
\end{equation}

where $\bm{F}$ is the nonlinear system of equations being solved and $\bm{x}$ is the state vector of unknowns. The deflation algorithm works by attempting to find a set of distinct solutions $\{\bm{x}_1^*, \bm{x}_2^*, ...\}$. Suppose a solution is found $\bm{x}_1^*$ to the above equation (by application of Newton's method, or some other way such as minimization of the residual). To find more solutions, one may apply a deflation operator to modify the system being solved, instead solving

\begin{equation}
    M(\bm{x};\bm{x}_1^*)\bm{F}(\bm{x})=0
\end{equation}

where

\begin{equation}
    M(\bm{x};\bm{x}_1^*) = \left(\frac{1}{||\bm{x}-\bm{x}_1^*||^p_2}\right) + \sigma
\end{equation}

is the deflation operator with power and shift\footnote{\cite{farrell_deflation_2015} added the shift parameter, which helps stabilize the deflation algorithm, which was otherwise prone to divergence and instability as found by \cite{brow_deflation_1971} and \cite{Wilkinson1963}.} parameters $p>0$ and $\sigma>0$, respectively, and $||\cdot||_2$ is the Euclidean norm. Under modest regularity conditions \citep{farrell_deflation_2015,pentland_multiple_2025}, solving the deflated system ensures the nonlinear solver does not return the known solution $\bm{x}_1^*$ (from the same initial guess) but rather a distinct solution $\bm{x}_2^*$ (if the nonlinear solver converges, which is not always guaranteed). To find further solutions, the system can then be deflated again using known solutions $\{\bm{x}_1^*,...,\bm{x}_N^*\}$ by simply applying the deflation operator repeatedly:

\begin{equation}
    \prod_{i=1}^N M(\bm{x};\bm{x}_i^*)\bm{F}(\bm{x}) = 0
\end{equation}

With this deflation method, one may find multiple solutions to a given system of equations. In this work, it will be applied to find multiple solutions to various equilibrium problems in DESC, where the nonlinear system being solved is the ideal magneto-hydrodynamic (MHD) equilibrium force balance, and minimization techniques are applied to find the solutions to the unmodified and modified systems. 

It is important to note that one has a choice in which part of the state (e.g. the equilibrium boundary or axis shape, a coil's geometry or current) to use in the operator $M(\bm{x};\bm{x}_i^*)$. It can be useful to not always use the entire state in the deflation, especially if one has some intuition on what part of the state should be distinct. The part of state used in the deflation operator will be explicitly stated in each application in this paper.\\
For the values of $\sigma$ and $p$ used, prior works have reached varying conclusions as to the effect of varying the deflation hyperparameters on the results. \cite{farrell_deflation_2015} found that the choice of hyperparameters can affect how many distinct solutions are successfully found through deflation, and this choice varied across different nonlinear problems. \cite{pentland_multiple_2025} had similar conclusions, and chose hyperparameters for each individual problem based on which performed best. \cite{riley_deflation_2025}, on the other hand, reported no meaningful performance sensitivity with respect to hyperparameters or deflation operators in most problems investigated there. In this work the results were found to be relatively sensitive to the hyperparameters, especially to the shift parameter (in the context of deflated equilibrium solves, \Autoref{sec:finding-multiple-solns-equilibrium} and \Autoref{sec:helical-core}), with larger shift parameters resulting in finding solutions which had better equilibrium force balance errors and were less prone to being pathological minima (such as having very sharp or self-intersecting surfaces). \textcolor{black}{As such, the value of $\sigma$ used in each section is the one found to provide the best performance for that specific problem, and will be stated explicitly. The $p$ value will be set to $2$ for the entirety of this work.}

\subsection{Application to Optimization Problems with Multiple Minima}

The deflation method can also be applied to optimization \citep{tarek_simplifying_2022, papadopoulos_computing_2021, riley_deflation_2025}, where unless one is dealing with very specialized cases like convex functions, multiple local minima generally exist . Global optimization of nonconvex functions is an extremely difficult problem, and so local optimizers are still the tool of choice when facing nonconvex optimization problems \citep{nocedal1999numerical}. Due to their nature, local optimization methods can only find local minima, and the found minima depend on the initial point (as well as the hyperparameters of the given optimization method). This is an analogous situation to the one in which the deflation method has already been applied above. However, in the optimization case it can also be useful to formulate the deflation operator as a nonlinear constraint, instead of using it to directly modify the objective function being minimized. To be clearer, we deal with the general optimization problem 

\begin{align*}
    &\min_x ~g(\bm{x})\\
    & s.t.~ c_i(\bm{x})=0 ~~, ~~ d_j(\bm{x})\leq0
\end{align*}

where $g(\bm{x})$ is the objective function, $c_i(\bm{x})=0$ are equality constraints and $d_j(\bm{x})$ are inequality constraints. In the formulation of \cite{tarek_simplifying_2022}, the deflation operator is added as an additional (nonlinear) inequality constraint defined as $M(\bm{x};\bm{x_1}) \leq r$ where $r$ is some chosen maximum value for $M(\bm{x};\bm{x_1})$, and is equivalent to constraining the solution to lie outside of some region around the known solution $\bm{x_1}^*$, since the deflation constraint can be written as:

\begin{equation}
    ||\bm{x}-\bm{x}_1^*||_2^p \geq \frac{1}{r-\sigma}
\end{equation}

This form of deflation will be used in this work to find multiple minima to various stellarator optimization problems in DESC, specifically stage-one and stage-two optimization, as it is easily added to existing optimization codes by simply augmenting the objective (to add a penalty term enforcing this constraint) or adding to the nonlinear constraints. Other deflation methods tailored for nonlinear least squares problems entail modifying the optimization algorithm itself, as is done in \cite{riley_deflation_2025} and was shown to increase robustness, the application of which to stellarator optimization will be the subject of future work.

\section{Finding Multiple Solutions to non-axisymmetric Ideal MHD Equilibrium Problems}\label{sec:finding-multiple-solns-equilibrium}

\color{black}
\subsection{The Fixed-Boundary and NAE-Constrained Equilibrium Problems}
The so-called "fixed-boundary" inverse-equilibrium problem is an approach to solve for nested-surface ideal MHD equilibrium equilibrium force balance, given for current density $\mathbf{J}$, magnetic field $\mathbf{B}$ and pressure $p$ as:

\begin{equation}\label{eqn:mhd-force-balance}
    \mathbf{F}=\mathbf{J}\times\mathbf{B}-\nabla p.
\end{equation}
In the fixed-boundary approach to inverse-equilibrium, the last-closed-flux-surface of the stellarator is prescribed, along with the net enclosed toroidal magnetic flux and the pressure and enclosed net toroidal current (both as flux functions). The boundary is typically given parametrically as a Fourier series for $R_b$ and $Z_b$ as a function of a poloidal angle $\theta$ and a toroidal angle $\zeta$, with $\zeta=\phi$ the cylindrical toroidal angle taken for this work. The geometry of the interior flux surfaces are then the degree of freedom of the problem, along with the poloidal stream function $\lambda$ which converts the poloidal angle used to describe the boundary into a straight-field-line poloidal angle. The problem is then to find the interior flux surface geometry which satisfies the ideal MHD equilibrium equations.

In the fixed-boundary approach as implemented in \texttt{DESC}, the equilibrium is sought by minimizing the ideal MHD equilibrium force balance error evaluated at collocation nodes as a function of the degrees of freedom of the problem (denoted as $\mathbf{f}(\mathbf{x})$) where in \texttt{DESC} the degrees of freedom are the Fourier-Zernike coefficients describing $R,Z, \text{and } \lambda$, $\mathbf{x}=[R_{lmn},Z_{lmn},\lambda_{lmn}]$ \citep{panici_desc_2023}. The prescriptions of the fixed-boundary problem are treated as constraints, with the fixed-boundary constraint translating to a simple linear constraint on the Fourier-Zernike coefficients to match the boundary coefficients:

\begin{equation}
    A \mathbf{x} = \mathbf{b}
\end{equation}

Where $A$ encapsulates the linear constraints and $b$ is the vector of boundary Fourier coefficients.\par

The near-axis-expansion (NAE) theory \citep{mercier_equilibrium_1964,Solovev1970,lortz1976equilibrium,garren_magnetic_1991,landreman_constructing_2019,rodriguez2024near} provides an asymptotic description of (typically omnigenous) stellarator equilibria by expanding the equilibrium force balance \autoref{eqn:mhd-force-balance} and omnigenity conditions order-by-order in powers of distance from the magnetic axis, resulting in a reduced set of equations compared to the full 3D equilibrium model. These equations may be solved orders of magnitude faster than the full MHD equilibrium equations, providing an efficient method of systematically exploring the stellarator configuration landscape. \par
However, due to the asymptotic nature of the NAE, the solution validity deteriorates with distance from the axis, and one typically must go to the full 3D MHD fixed-boundary equilibrium approach to construct solutions suitable for higher-fidelity analyses. The conventional method of solving fixed-boundary equilibria is at odds with the very nature of the NAE: the fixed-boundary approach requires from the NAE the outermost flux surface of the equilibrium, which is the least accurate part of the NAE solution, rendering this method of converting NAE solutions to full 3D MHD equilibrium solutions poor at low aspect ratios. The resulting full 3D MHD equilibrium solutions do not compare well to the NAE in terms of on-axis rotational transform and omnigenity quality at these low aspect ratios, ruining the desirable qualities encoded into the NAE solution. \par

\cite{panici_extending_2025} recently developed a new approach to connecting NAE solutions to full 3D MHD equilibria in \texttt{DESC} which can retain the desirable aspects of the NAE in the core of the equilibrium, even at low aspect ratios. The NAE considered in the work provides a solution to the inverse-equilibrium problem, so all of the information of the equilibrium is encoded into the geometry of the flux surfaces of its solution. For example, a first-order NAE solution will, for a given magnetic axis $\mathbf{r}_0(\phi)$, yield the first-order (in normalized square root of toroidal flux, $\rho$) flux surface geometry:

\begin{equation}
    \mathbf{r}(\rho, \theta,\phi) = \mathbf{r}_0(\phi) + \rho R_1(\theta,\phi)\hat{\mathbf{R}}  + \rho Z_1(\theta,\phi)\hat{\mathbf{Z}}.
\end{equation}

Where

\begin{align}
R_1(\theta,\phi) &= R_{1c,n} \cos(\theta)\mathcal{F}_n(\phi) + R_{1s,n} \sin(\theta)\mathcal{F}_n(\phi)\\
Z_1(\theta,\phi) &= Z_{1c,n} \cos(\theta)\mathcal{F}_n(\phi) + Z_{1s,n} \sin(\theta)\mathcal{F}_n(\phi)\\
\mathcal{F}_n(\phi) &= \begin{cases} cos(|n|N_{FP}\phi) \text{for } n\geq 0 \\
sin(|n|N_{FP}\phi) \text{for } n < 0 
\end{cases}
\end{align}

is the first-order NAE information encoded as Fourier coefficients for $R$ and $Z$ in the poloidal and toroidal angles $\theta$ and $\phi$ (see \cite{panici_extending_2025} for more details). The first-order NAE-constrained equilibrium problem is then defined as solving for ideal MHD equilibrium subject to prescribed constraints on the enclosed toroidal flux, flux functions of pressure and net enclosed toroidal current, as well as the new constraint that the equilibrium geometry asymptotically match that of the NAE to a given order in $\rho$. This constraint can be written as 

\begin{equation}
    A_{NAE} \mathbf{x} = \mathbf{b}_{NAE}
\end{equation}

where now $A_{NAE}$ encodes the linear constraints that match the \texttt{DESC} equilibrium's flux surface geometry to the NAE behavior $\mathbf{b}_{NAE}$ to first-order in $\rho$ (higher orders are possible \citep{panici_extending_2025}, first-order was chosen for simplicity in this work). Note that there is no constraint on the last-closed-flux-surface in this formulation: the boundary is allowed to vary as part of the equilibrium procedure in order to minimize force balance.\par

If one compares numbers of constraints imposed on the equilibrium, one sees that the conventional fixed-boundary approach applies $2(2M+1)(2N+1)$ constraints on the problem (where $M$ and $N$ are the maximum poloidal and toroidal resolution of the equilibrium), while the first-order NAE-constraint approach only imposes $12N + 6$. This drastically lower number of constraints for the NAE-constrained problem implies that there is potentially more than one global MHD solution consistent with a given set of NAE constraints, which deflation is well-suited to investigating.

\color{black}

\subsection{Finding Multiple Solutions to the NAE-constrained Equilibrium Problem}\label{sec:nae-multiple-solns}

One can consider the NAE-constrained equilibrium problem as the following system to be solved:

\begin{align} \label{eq:nae-equil-prob-statement}
    \bm{f}(\bm{x}) &= 0 ~ \text{ s.t. } A_{NAE}\bm{x}=\bm{b}_{NAE}
\end{align}

where $A_{NAE}$ and $\bm{b}_{NAE}$ encapsulate the linear NAE-constraints on the Fourier-Zernike coefficients $\bm{x}$ which describe the equilibrium solution, and $\bm{f}$ is the ideal MHD equilibrium force balance error evaluated at collocation nodes \citep{panici_desc_2023}. Like any inverse equilibrium solver, \texttt{DESC} must begin from some initial coordinate mapping $\bm{x}_0$, from which the minimization procedure will proceed to produce an ideal MHD equilibrium solution $\bm{x}^*$ s.t. $\bm{f}(\bm{x}^*)=0$. In reality, due to the general non-existence of non-axisymmetric fields with nested flux surfaces, as well as the finite numerical resolutions employed, $\bm{f}(\bm{x}^*)\neq0$, but rather $\approx0$, but this caveat is not an issue here. One simply seeks equilibria in the usual way in DESC, as local minima of the system of residuals being minimized.

The initial coordinate mapping $\bm{x}_0$ will be taken as the Fourier-Zernike fit of the flux surface geometry evaluated from the NAE (obtained using pyQSC or pyQIC). The NAE-constrained equilibrium problem \ref{eq:nae-equil-prob-statement} is then solved as usual in order to obtain the first solution $\bm{x}_1^*$. To find further solutions, one then solves the deflated equilibrium problem:

\begin{align} \label{eq:deflated-nae-equil-prob-statement}
    \prod_iM(\bm{x};\bm{x}_i^*)\bm{f}(\bm{x}) &= 0 ~ \text{ s.t. } A_{NAE}\bm{x}=\bm{b}_{NAE}
\end{align}

    where the deflation operator $M$ ensures that, if a solution is found, it will be distinct from the past solutions. Because of the caveat that solutions $\bm{f}=0$ are not typically found but rather $\bm{f}\approx0$, once a solution $\bm{x}_i$ to \autoref{eq:deflated-nae-equil-prob-statement} is found, that solution is used as an initial guess and the non-deflated problem \autoref{eq:nae-equil-prob-statement} is solved to yield the solution $\bm{x}_i^*$. This is done to ensure that $\bm{x}_i^*$ is actually a solution to the NAE-constrained equilibrium problem, and not merely some local minimum of the deflated equilibrium problem. So effectively, the deflated solve is done only to find an initial condition near a new local minimum of \autoref{eq:nae-equil-prob-statement}. This is summarized in Algorithm \autoref{alg:nae-deflation}. In this section, the deflation operator was applied to the boundary Fourier coefficients of the solution, e.g. on $R^b_{mn}, Z^b_{mn}$, and the power and shift parameters used were $p=2$ and $\sigma=500$. These power and shift were chosen as they were found to work best after performing a small parameter scan. \textcolor{black}{Computationally speaking, having even powers may be slightly advantageous over odd as they then avoid instantiating the square-root operation, which can be more expensive (either in computational cost or in having a more complex automatic-differentiation derivative graph) than the case of an even $p$ where the square roots are avoided.} Having a larger shift was found to especially be important for finding equilibria which did not suffer from poor force balance due to the deflation cost being dominant over the equilibrium force cost, \textcolor{black}{which was observed to occur at small $\sigma$ values in this section. The failure state which was observed to occur at too low of a shift value was that the deflation cost dominated over the equilibrium, and the resulting new state from which to solve for equilibrium is no longer in any sensible region of the phase space (for example, extreme elongation, or nearly-intersecting flux surfaces). Thus the subsequent equilibrium solve results in a poor equilibrium (as determined by too-high force balance error). Conversely, a too-high value of sigma would have the opposite problem: it would not penalize the deflated equilibrium solve enough to make it move any significant distance away from the already-found equilibrium state, and so the subsequent equilibrium solve (with the deflation constraint relaxed) would just return to the already-found equilibrium.}

\begin{algorithm}
\caption{NAE-Constrained Equilibrium Deflation}\label{alg:nae-deflation}
\begin{algorithmic}[] 
    \Require $\bm{x}_0$ initial coordinate mapping, $N_{defl}$ number of deflation iterations, $p$ power parameter, $\sigma$ shift parameter
    \State{$\rightarrow$Solve NAE-Constrained Equilibrium}
    \State $\bm{x_1^{NAE}} = \argmin_{x} \bm{f}(\bm{x})=0~~\text{s.t}~~A_{NAE}\bm{x}=\bm{b}_{NAE}~~\text{from initial state}~~\bm{x}_0$ 
    \State{$\rightarrow$Solve Fixed-Boundary Equilibrium}
    \State $\bm{x_1^*} = \argmin_{x} \bm{f}(\bm{x})=0~~\text{s.t}~~A\bm{x}=\bm{b} ~~\text{from initial state}~~\bm{x}^{NAE}_1$ 
    
    \State $i = 1$
    \State{$\rightarrow$Initialize list of solved equilibria to deflate with}
    \State eqs=$[\bm{x}_1^*]$ 
    \While{$i < N_{defl}$}
    \State{$\rightarrow$Solve Deflated NAE-Constrained Equilibrium}
    \State $\bm{x_i} = \argmin_{x} \prod_iM(\bm{x};eqs)\bm{f}(\bm{x})=0~~\text{s.t}~~A_{NAE}\bm{x}=\bm{b}_{NAE}~~\text{from initial state}~~\bm{x}_0$ 
    \State{$\rightarrow$Solve NAE-Constrained Equilibrium}
    \State $\bm{x_i^{NAE}} = \argmin_{x} \bm{f}(\bm{x})=0~~\text{s.t}~~A_{NAE}\bm{x}=\bm{b}_{NAE}~~\text{from initial state}~~\bm{x}_i$ 
    \State{$\rightarrow$Solve Fixed-Boundary Equilibrium}
    \State $\bm{x_i^*} = \argmin_{x} \bm{f}(\bm{x})=0~~\text{s.t}~~A\bm{x}=\bm{b}~~\text{from initial state}~~\bm{x}_i^{NAE}$ 
    \State eqs.append($\bm{x}_i^*$)
    \State $i=i+1$
    \EndWhile
    \State \textbf{return} eqs
\end{algorithmic}
\end{algorithm}

\begin{figure}
    \centering
    \includegraphics[width=1.1\linewidth]{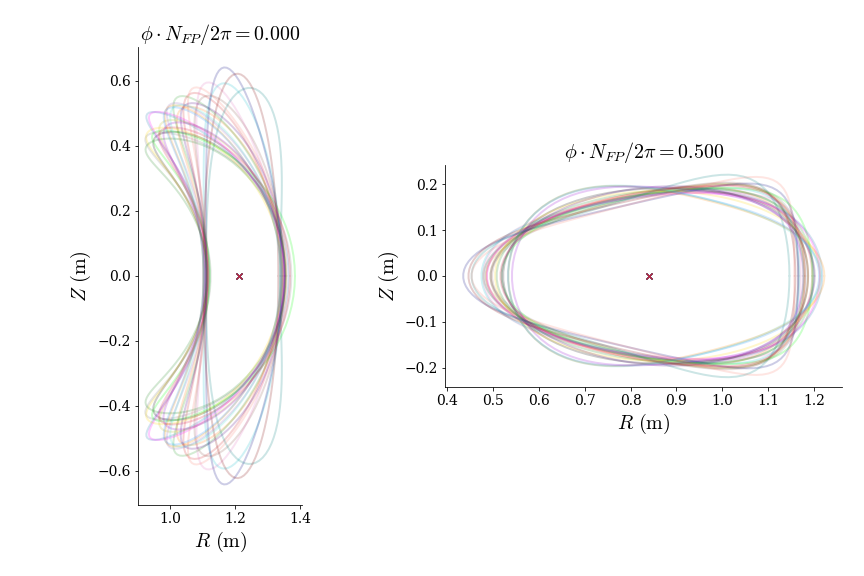}
    \caption{Boundaries at two different toroidal angles for the 25 deflated first-order NAE-constrained equilibria.}
    \label{fig:QA-NAE-deflated-bdries}
\end{figure}

\begin{figure}
    \centering
    
    \hspace{-0.75cm}
    \begin{subfigure}[b]{0.525\textwidth}
        \centering
        \includegraphics[width=\linewidth]{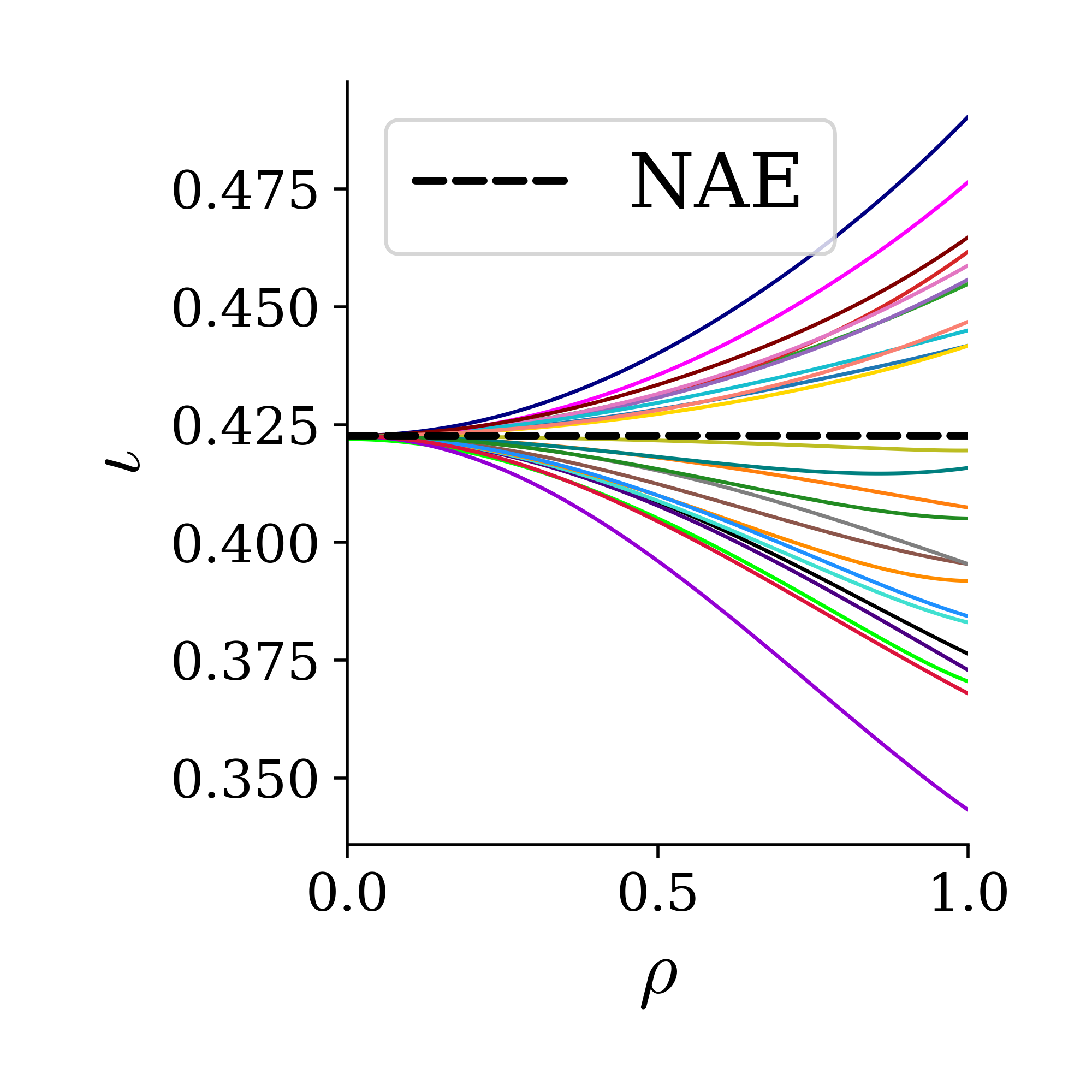}
        \caption{Rotational transform profiles}
    \label{fig:QA-NAE-deflated-iotas}
    \end{subfigure}
   \hspace{-0.75cm} 
    \begin{subfigure}[b]{0.55\textwidth}
        \centering
        \includegraphics[width=\linewidth]{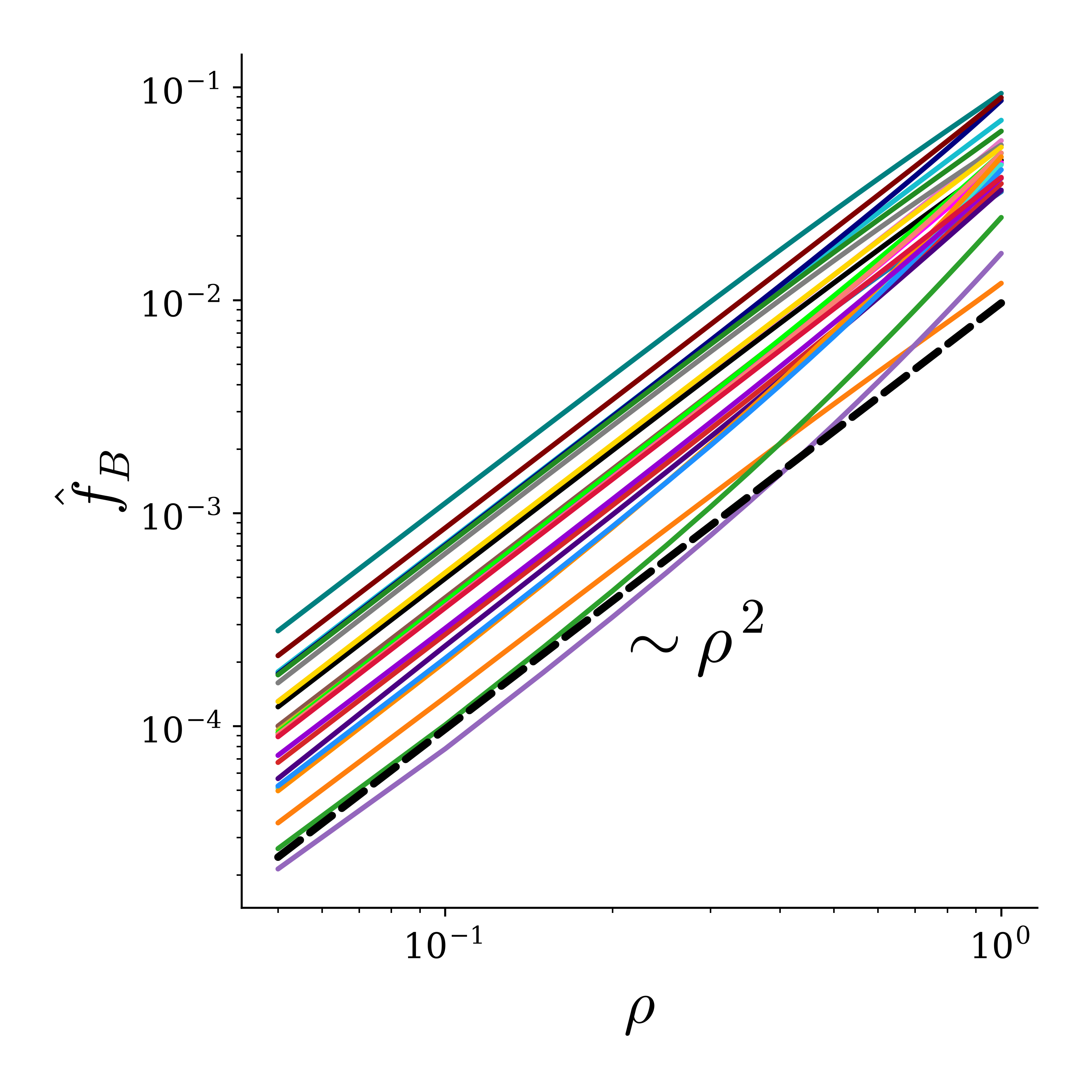}
        \caption{QS error}
        \label{fig:QA-NAE-deflated-QS-errs}
    \end{subfigure}
    \caption{(a) Rotational transform profiles and  (b) QS error for the 25 deflated first-order NAE-constrained equilibria.}
\end{figure}

\begin{figure}
    \hspace{1.5cm}
    \includegraphics[width=0.95\linewidth]{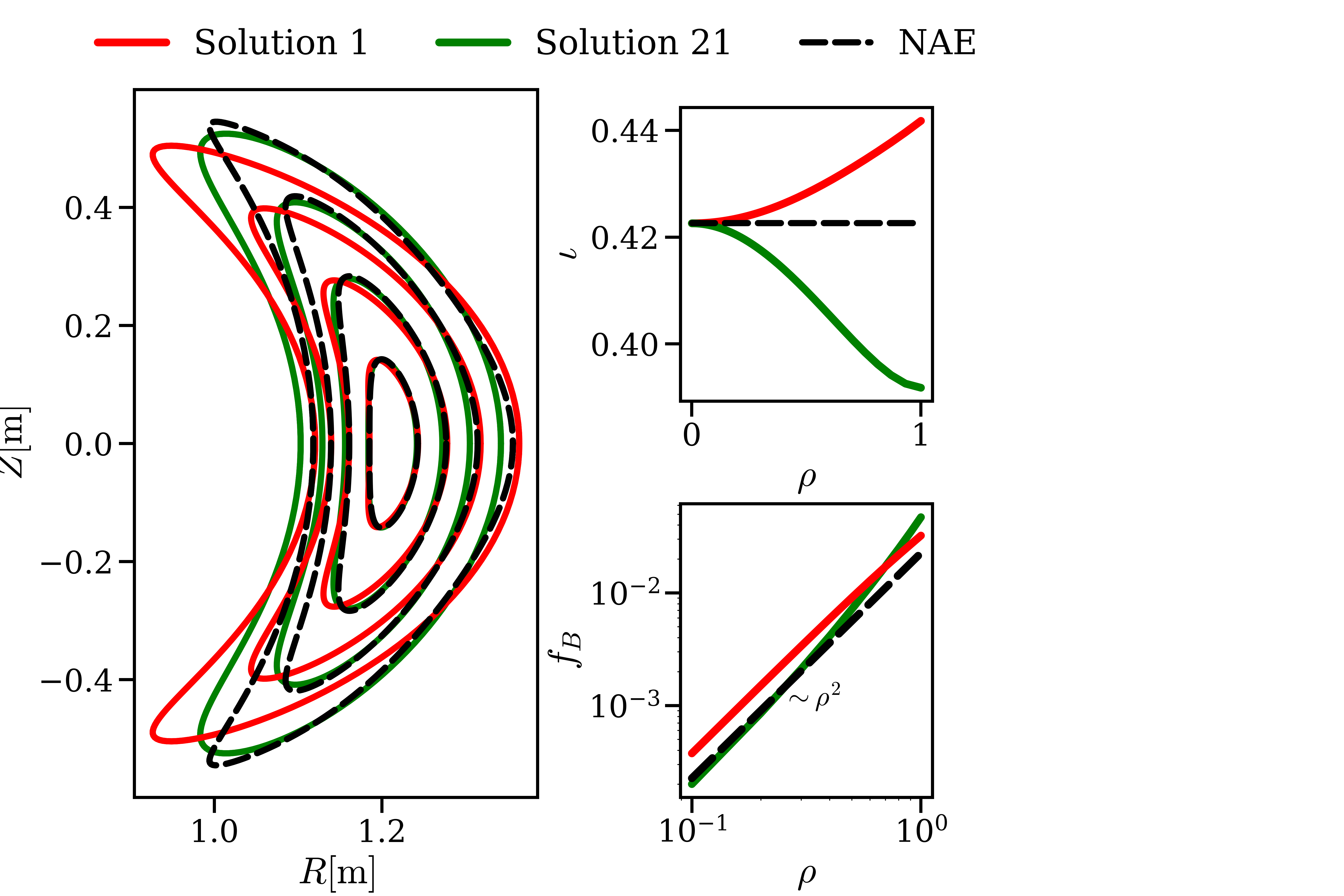}
    \caption{A comparison of two of the 25 deflated first-order NAE-constrained equilibria. (Left) The flux surfaces of the two equilibria at the $\phi=0$ cross-section, compared to the original NAE surfaces. (Right top) The rotational transform profile for the two solutions, with both matching well near-axis. (Right Bottom) The Boozer QS error metric for the two solutions, with both displaying the expected quadratic scaling of the QS error with $\rho$.}
    \label{fig:QA-NAE-deflated-two-comparison}
\end{figure}

To provide an example of a successful application of deflation to this problem, we take the first-order quasi-axisymmetric (QA) NAE solution from \cite{panici_extending_2025}, \textcolor{black}{a near-axis construction with on-axis field of $B_0\approx 1$T  based off of the vacuum two-field-period "precise QA" equilibrium from \cite{landreman_magnetic_2022}}. We evaluate the NAE surfaces at $r=0.25m$ (for an aspect ratio of $\sim 4$), \textcolor{black}{and fit the resulting flux surface geometry with a Fourier-Zernike basis in \texttt{DESC} with radial, poloidal, and toroidal resolution of $L=12, M=4, N=8$, respectively, resulting in our initial state which we denote $\mathbf{x}_0$.}  We follow the above procedure, first finding an NAE-constrained solution \textcolor{black}{($\mathbf{x}_{1}^{NAE})$}, then performing deflated NAE-constrained solves to find initial conditions \textcolor{black}{$\mathbf{x_i}$} from which to then find new NAE-constrained solutions. We apply this procedure 25 times to result in 25 distinct NAE-constrained solutions. Like in \cite{panici_extending_2025}, after each NAE-constrained equilibrium solve, we then solve a conventional fixed-boundary equilibrium solve to ensure equilibrium in the traditional sense \textcolor{black}{(we denote the fixed-boundary-solved equilibria $\mathbf{x_i^*}$)}, and we note that every presented solution here has a volume averaged force balance error below one percent \textcolor{black}{(In fact, all have on the order of 0.01\% or less normalized error, see \autoref{app:nae-soln-plots})}. The NAE-constraints used in this section were on the first-order $R,Z$ flux surface geometries of the equilibrium. \autoref{fig:QA-NAE-deflated-bdries} show the boundary shapes and magnetic axis positions for each of the found NAE-constrained solutions at two toroidal angles, while Figures \ref{fig:QA-NAE-deflated-iotas} and \ref{fig:QA-NAE-deflated-QS-errs} shows the rotational transform profiles and Boozer QS errors ($\hat{f}_B$ \citep{rodriguez_measures_2022, dudt_desc_2023}) for each solution. It can be seen that a wide range of both geometries and rotational transform profiles were found, all of which share the same axis position and on-axis rotational transform, as well as a general adherence to the expected scaling of the QS error with $\rho^2$. To look more closely at the different solutions found, \autoref{fig:QA-NAE-deflated-two-comparison} shows the flux surfaces, rotational transform profiles, and QS error for two solutions, the first (i.e. undeflated) and the 21st solutions found. While both share the same magnetic axis, on-axis rotational transform, and both are QS to first-order in $\rho$, the sign and strength of the shear is different, as well as the outer flux surface geometries of the two equilibria. Both could serve as good initial guesses for further stage-one optimizations if one desired a QA equilibrium, with each being individually useful if one desires a different rotational transform profile (the second solution being a good point from which to optimize for larger shear, for example). \autoref{app:nae-soln-plots} shows some further metrics of these QA solutions, emphasizing the wide range of possible solutions which match similar core properties, including some which have stable magnetic wells (such as the 21st solution).

It is worth noting that when performing deflated solves, the only guarantee is that if a solution is found, the obtained state parameter values do not exactly match that of the prior deflated solutions. However, depending on how the state is parametrized, the same physical solution may correspond to multiple different sets of state parameter values. As a concrete example, a stellarator-symmetric equilibrium with $N_{FP}$-symmetry rotated around toroidally through the angle $\pi/N_{FP}$ can be represented by the same stellarator-symmetric basis and appear different to the deflation operator, but have the same exact physical properties. This is not an issue for the NAE-constrained equilibria, as the fixing of the magnetic axis removes this possible degeneracy, but it will have implications for stage one and stage two optimizations, as discussed later in \Autoref{sec:finding-multiple-solns-optimization}.

\subsection{Helical Core Equilibria}\label{sec:helical-core}

In work first carried out by \cite{cooper_tokamak_2010}, it was found that for certain profiles and axisymmetric boundaries, it was possible for \textcolor{black}{3D} ideal MHD fixed-boundary equilibrium solvers to find two distinct equilibria depending on the initial guess: one equilibrium solution with a wholly axisymmetric interior \textcolor{black}{(despite the allowed freedom in the non-axisymmetric modes of the geometry)} and one non-axisymmetric with a helically distorted magnetic axis, both having the same fixed axisymmetric boundary. \textcolor{black}{Since Cooper's pioneering work, many other studies have been conducted on the so-called "snake" or "bifurcated helical core" equilibrium states in tokamaks \citep{cooper_jet_2011, graves_magnetohydrodynamic_2012, cooper_bifurcated_2013, strumberger_mhd_2014, wingen_use_2018}, among which a study by \cite{rakha_shear_2019} showed that both axisymmetric and helical core solutions are numerically converged according to standard convergence criteria for VMEC. These studies also found that the axisymmetric states tended to be either unstable or marginally stable to internal kink modes, potentially explaining the helical core as a saturated internal kink state.} Cooper \textcolor{black}{(and each of the later studies)} obtained the second, helical core solution by initializing the equilibrium code (VMEC \textcolor{black}{ or ANIMEC}) coordinate mapping with a slightly perturbed magnetic axis, which then would converge to the helical core state when solved. However, it would be interesting to find such states without the use of judiciously perturbed initial states, as the necessary perturbation size is unknown a priori. To approach this, we state the typical fixed-boundary equilibrium problem in \texttt{DESC} as

\begin{align} \label{eq:fixed-bdry-equil-prob-statement}
    \bm{f}(\bm{x}) &= 0 ~ \text{ s.t. } A\bm{x}=\bm{b}
\end{align}

where here $A$ are the linear constraints on the Fourier-Zernike coefficients $\bm{x}$ which enforce the fixed-boundary geometric constraint. One may apply the deflation technique to this system:

\begin{align}\label{eq:deflated-fixed-bdry-equil-prob-statement}
    \prod_iM(\bm{x};\bm{x}_i^*)\bm{f}(\bm{x}) &= 0 ~ \text{ s.t. } A\bm{x}=\bm{b}
\end{align}
 where $\bm{x}_i^*$ are solutions to \autoref{eq:fixed-bdry-equil-prob-statement}. Like in the prior section, the procedure adopted is to first solve \autoref{eq:fixed-bdry-equil-prob-statement} to obtain $\bm{x}_1^*$ from some initial coordinate mapping $\bm{x}_0$, then use the same initial mapping $\bm{x}_0$ to solve the deflated equilibrium \autoref{eq:deflated-fixed-bdry-equil-prob-statement} to obtain $\bm{x}_2$, which is then used as an initial condition from which to solve \autoref{eq:fixed-bdry-equil-prob-statement} to obtain $\bm{x}_2^*$. In this problem, because the desired additional distinct solution is one with a different magnetic axis, the deflation operator was applied only to the magnetic axis position, i.e. if the current state's magnetic axis geometry is $\bm{x}_{axis}$ and the solution axis $\bm{x}_{axis}^*$, the deflation operator is $M(\bm{x}_{axis};\bm{x}_{axis}^*)$. The axis position is represented as the toroidal Fourier series describing the curve, so $\bm{x}_{axis} = [R_n, Z_n]$ where 
 
 \begin{equation}
     \begin{aligned}
         R_{axis}(\phi) &= R_n\mathcal{F}_n(\phi)\\
         Z_{axis}(\phi) &= Z_n\mathcal{F}_n(\phi)\\
         \mathcal{F}_n(\phi) &= \begin{cases}
             cos(nN_{FP}\phi) ~ n\geq0\\
             sin(nN_{FP}\phi) ~ n<0\\
         \end{cases} \\
     \end{aligned}
 \end{equation}
The power and shift parameters used were $p=2$ and $\sigma=100$, again chosen as the larger shift was found to result in a more stable deflated equilibrium solve and better final equilibrium.

This procedure was applied to an equilibrium described in \cite{cooper_tokamak_2010}. The axisymmetric fixed-boundary shape, pressure and rotational transform profiles are given by

 \begin{equation}
     \begin{aligned}
         R(\theta) &= 0.8 + 0.2cos\theta + 0.06cos2\theta~ [m]\\
         Z(\theta) &= 0.48sin\theta~ [m]\\
         p(\rho) &= p_0(1-\rho^2)\\
         \iota(\rho)&= 0.9 + 0.2\rho^2 - 0.8\rho^{12}
     \end{aligned}
 \end{equation}
     
\begin{figure}
    \centering
    
    \hspace{-2cm}
    \begin{subfigure}[b]{0.6\textwidth}%
        \centering%
        \includegraphics[width=\linewidth]{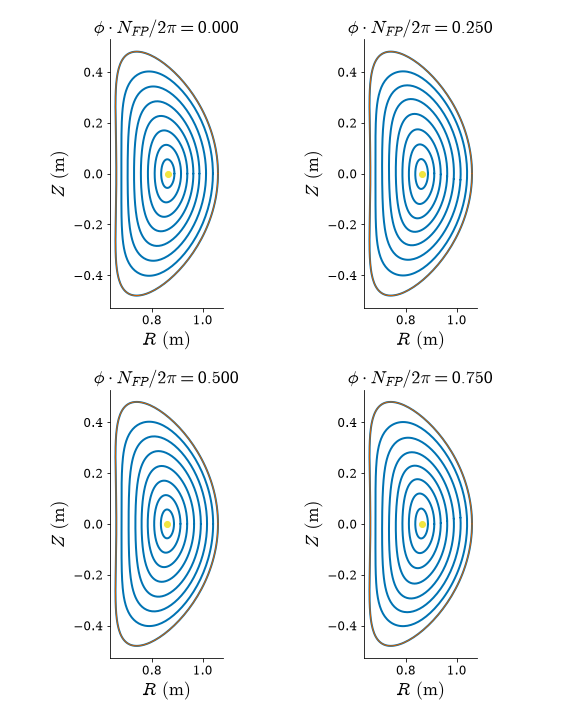}%
        \caption{Axisymmetric equilibrium solution.}%
        \label{fig:left}%
    \end{subfigure}
    \hspace{-1cm}
    \begin{subfigure}[b]{0.6\textwidth}%
        \centering%
        \includegraphics[width=\linewidth]{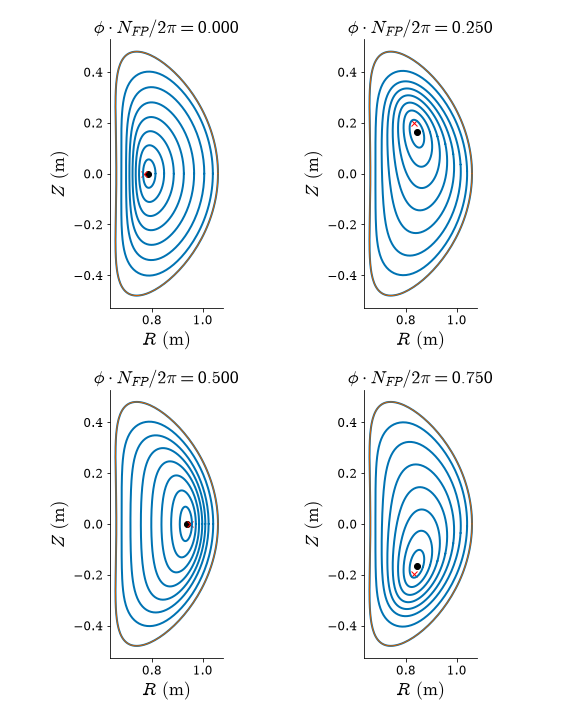}%
        \caption{Helical core equilibrium solution.}%
        \label{fig:right}%
    \end{subfigure}%
    
    \caption{The two branches of equilibrium found for the helical core state presented in \cite{cooper_tokamak_2010}. (Left) The axisymmetric equilibrium state, found when solving from an axisymmetric initial condition. (Right) The helical core equilibrium, found when solving from the deflated equilibrium solve as the initial state. The solved equilibrium's axis is shown as a black dot, while the deflated solve solution's is a red cross, showing that the deflated equilibrium problem encouraged an axis perturbation in the correct direction.
    }
    \label{fig:helical-core}
\end{figure}

Where $p_0=49,700~\text{Pa}$ was chosen to yield a volume-averaged plasma $\beta\approx0.5\%$, and the net enclosed toroidal flux is $1$ Weber. \autoref{fig:helical-core} shows the flux surfaces of the axisymmetric equilibrium state $\bm{x}_1^*$ on the left, and the non-axisymmetric helical core branch $\bm{x}_2^*$ on the right. Each have a volume-averaged force balance error less than 1\%. The right figure shows also in red the axis of the result of the deflated equilibrium solve, $\bm{x}_2$, where the deflated equilibrium solve results in an axis perturbation similar to what was found to be necessary in \cite{cooper_tokamak_2010}. It is worth reiterating this point: the helical axis state was found without any prescient choice of perturbation to the axisymmetric initial condition. Instead, the deflated equilibrium solve found the second initial condition $\bm{x}_2$, from which the equilibrium solver then converged to the helical core equilibrium $\bm{x}_2^*$. This shows the promise of deflation methods in finding effective initial conditions for cases where one may suspect multiple equilibrium states, and also warrants further search for other cases where such potential neighboring states exist.

\section{Deflation for finding multiple local minima to nonconvex optimizations}\label{sec:finding-multiple-solns-optimization}

\subsection{Stage one}\label{sec:stage-one-deflation}
Stage-one stellarator optimization almost always have multiple local minima due to the complex, multi-objective landscapes typically explored. These multiple objectives may compete with each other \citep{bindel_understanding_2023,laia_data-driven_2026}, and nonlinear constraints are usually enforced with penalty methods in combined sum of squares objectives. Constrained optimization methods have been shown to improve the situation somewhat \citep{conlin_stellarator_2024}, but even these methods can still lead to poor local minima. Conventionally, the only recourse left is to launch many optimizations beginning from many initial conditions, or resort to global optimization methods, both of which are expensive approaches. Instead, one may apply deflation to attempt to find distinct local minima in the optimization landscape, starting from the same initial guess \citep{tarek_simplifying_2022}.  
In this section we employ deflation to find multiple local minima to a quasi-helically symmetric (QH) stellarator optimization problem. Constrained optimization using the augmented Lagrangian approach will be employed, and the deflation will be in the form of an additional nonlinear constraint.

To state the specific problem clearly, we will seek a quasi-helically symmetric vacuum stellarator. The optimization will target the two-term quasisymmetry (QS) metric \citep{helander_intrinsic_2008, landreman_magnetic_2022, rodriguez_measures_2022}, while constraining the aspect ratio, rotational transform, and ideal MHD force balance through. The two-term QS metric is given by $f_{QS}=\left[(M\iota - N)(\bm{B}\times\nabla\psi))\cdot\nabla B - (MG+\iota I)\bm{B}\cdot\nabla B\right]^2$, with $\bm{B}$ being the magnetic field and $B$ its magnitude, $(M,N)=(1,4)$ being the QS helicity, $\psi$ being the normalized toroidal flux, $\iota$ the rotational transform and $I,G$ being the Boozer toroidal and poloidal currents, respectively. We fix the major radius of the equilibrium $R_0$, aspect ratio $R_0/a$, net enclosed toroidal flux $\Psi$ and vacuum net toroidal current $I_{tor}(\rho)$ and pressure profile $p(\rho)$ during the solve. We then minimize this objective while constraining the average rotational transform $\bar{\iota}$ to be between $0.7$ and $0.3$, as well as enforcing ideal MHD force balance. The constraints for the undeflated optimization problem are then:

\begin{equation}
    \begin{aligned}
        R_0 &= 1~\text{m}\\
        R_0/a &= 8\\
        0.7 < &\bar{\iota} < 3\\
        p(\rho)&=0\\
        I_{tor}(\rho)&=0\\
        \Psi &= 0.04 ~\text{Tm}^2\\
        \bm{J}\times\bm{B} - \nabla p &= 0
    \end{aligned}
\end{equation}

\begin{figure}
    \centering
    \includegraphics[width=1.05\linewidth]{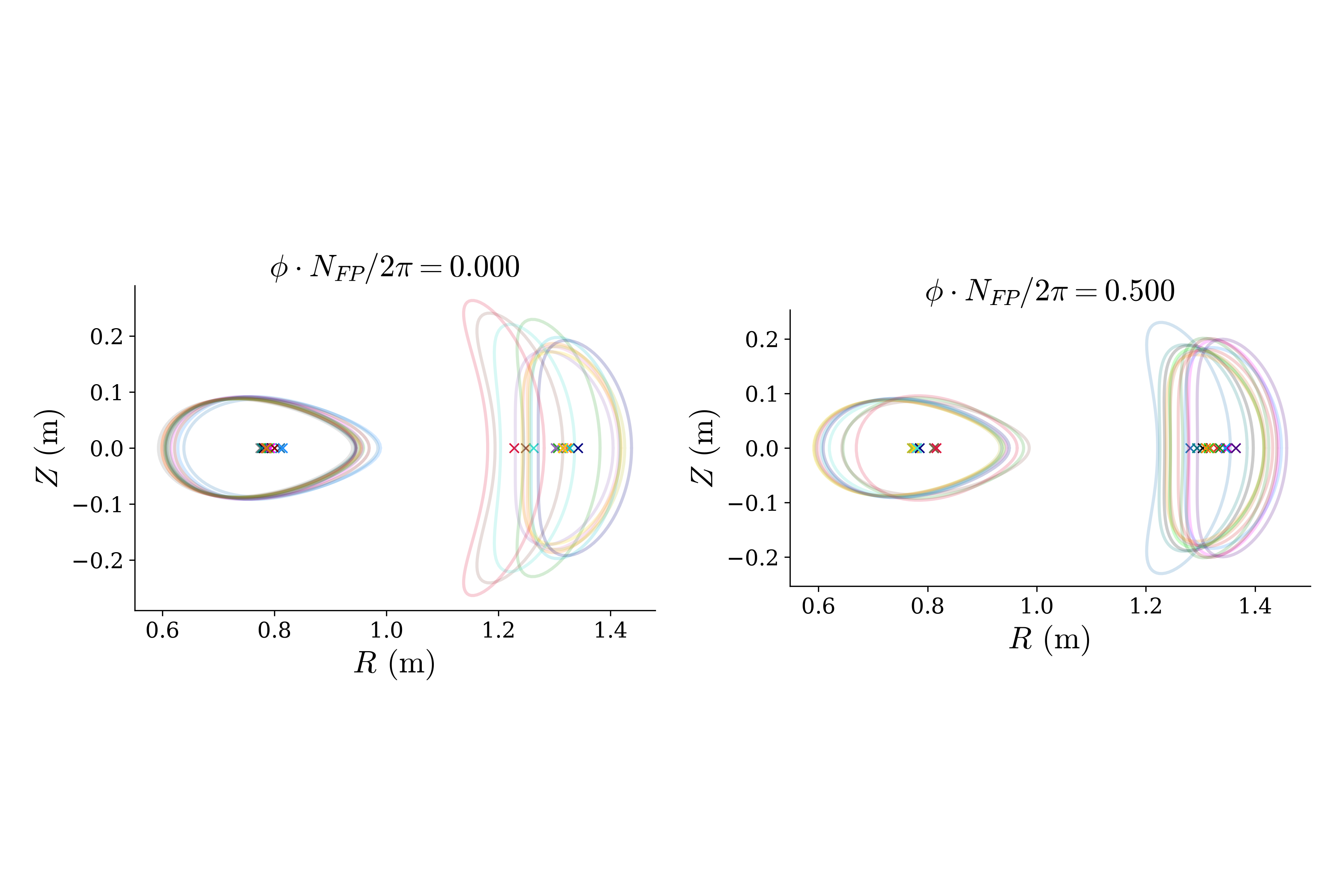}
    \caption{Boundaries at two different toroidal angles for the 24 optimized QH equilibria found through deflation which passed the filters.}
    \label{fig:QH-deflated-bdries}
\end{figure}

\begin{figure}
    \centering
    
    \begin{subfigure}[b]{0.49\textwidth}
        \centering
        \includegraphics[width=\linewidth]{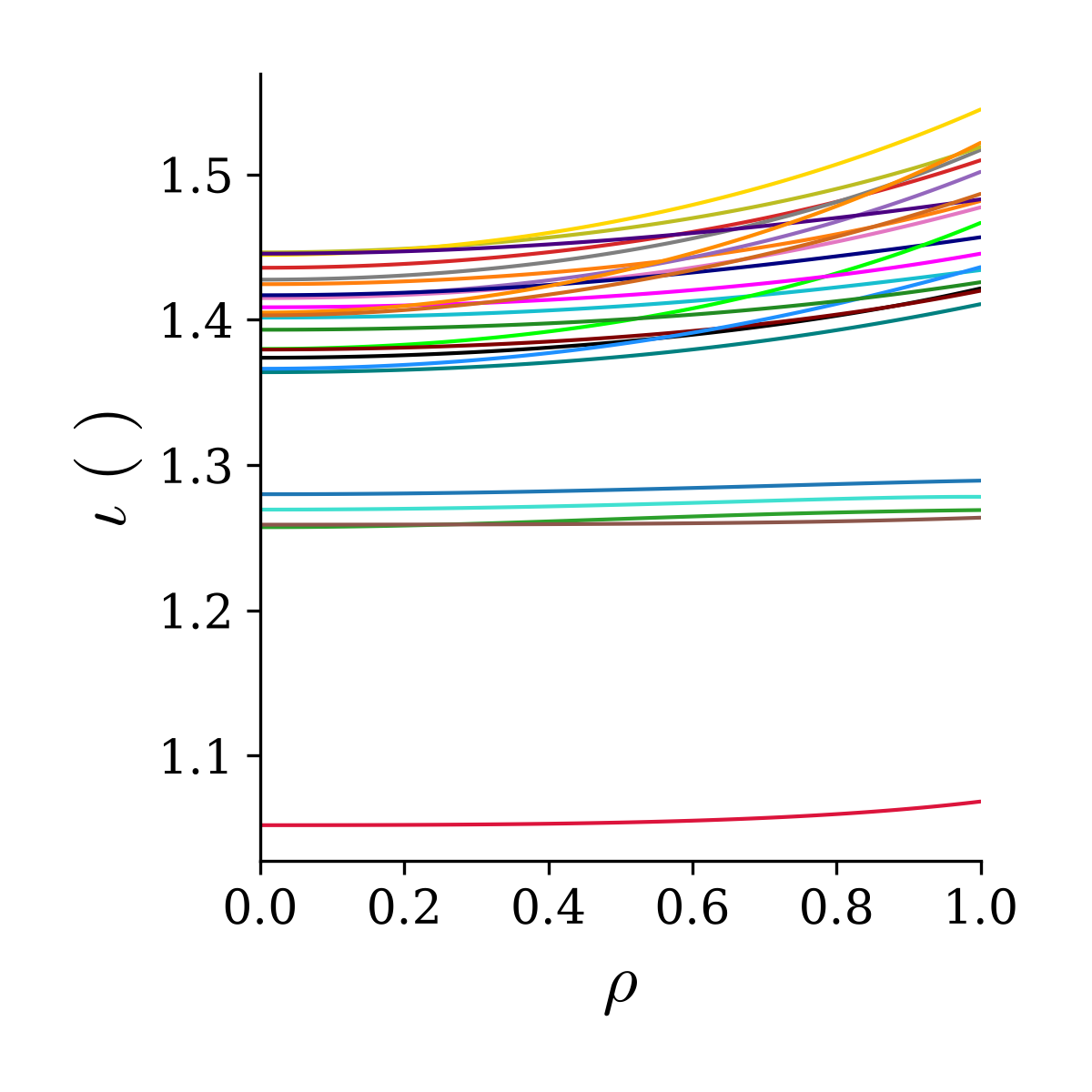}
        \caption{Rotational transform profiles}
    \label{fig:QH-deflated-iotas}
    \end{subfigure}
    \hfill 
    \begin{subfigure}[b]{0.49\textwidth}
        \centering
        \includegraphics[width=\linewidth]{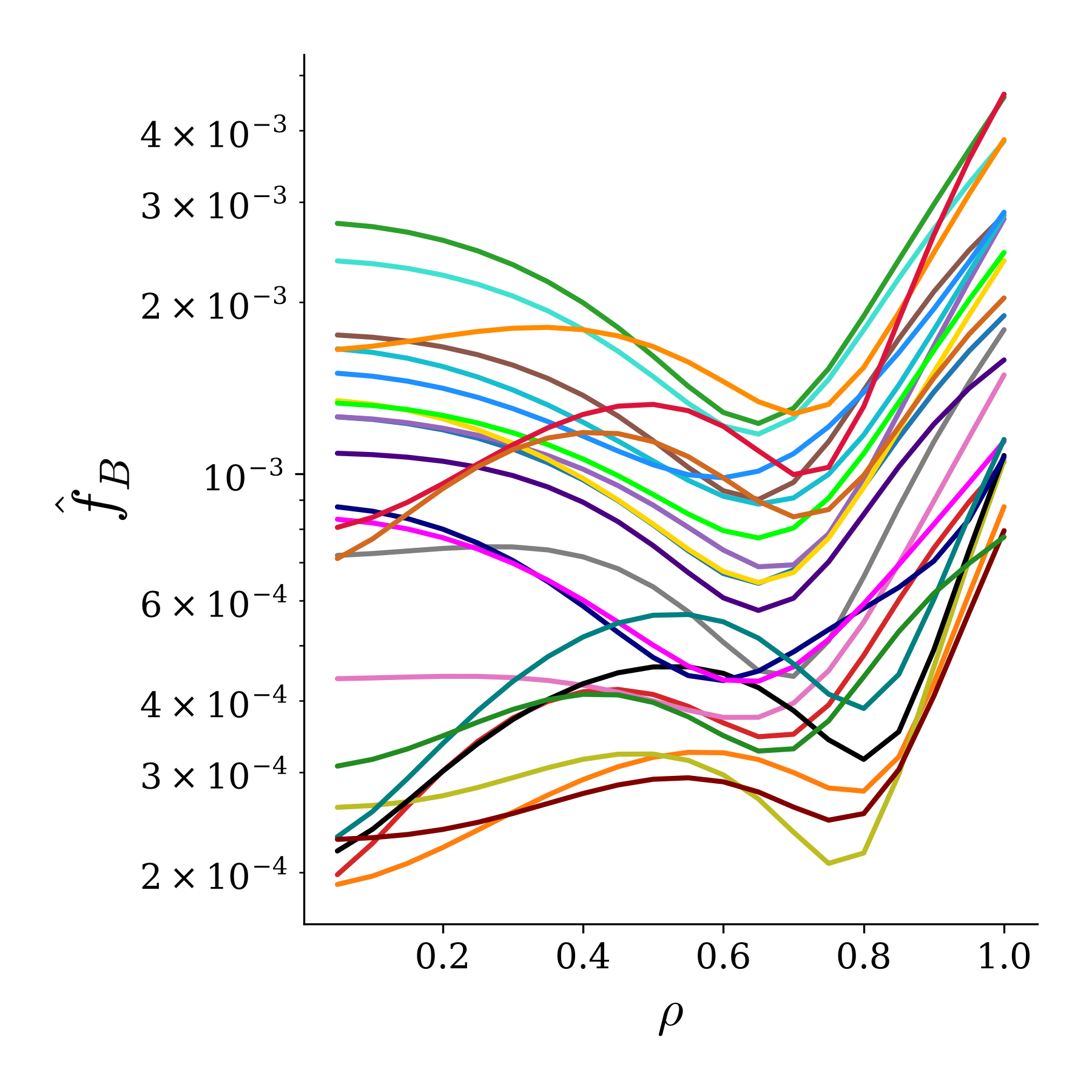}
        \caption{QS error}
        \label{fig:QH-deflated-QS-errs}
    \end{subfigure}
    \caption{Rotational transform profiles and QS error for the 24 optimized QH equilibria found through deflation which passed the filters.}
\end{figure}

\begin{figure}
    \centering
    \includegraphics[width=0.95\linewidth]{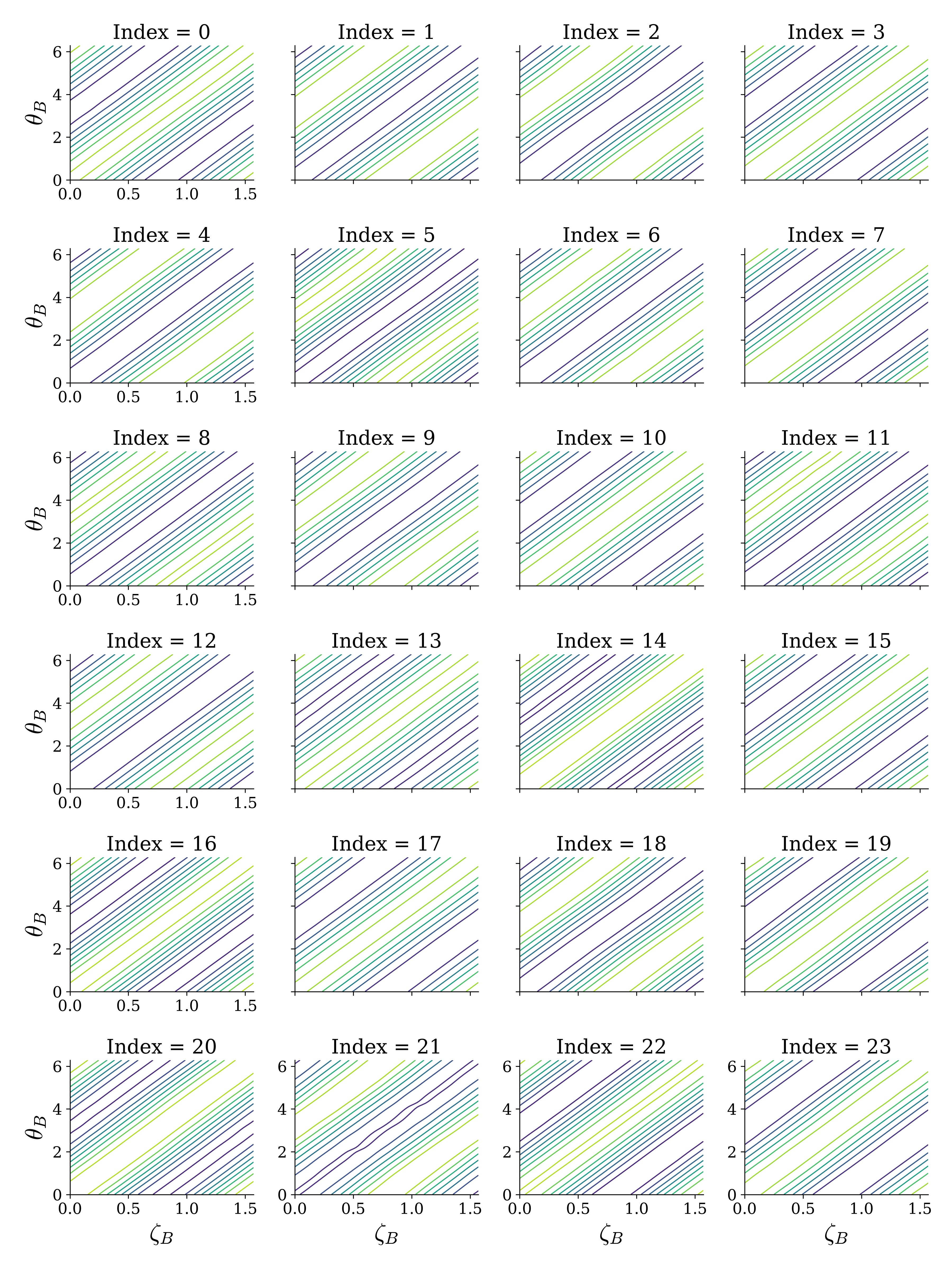}
    \caption{Magnetic field magnitude contours in Boozer coordinates on the boundary for the 24 optimized QH equilibria found through deflation which passed the filters.}
    \label{fig:QH-deflated-boozer-plots}
\end{figure}

This problem is initially solved using the augmented Lagrangian approach outlined in \cite{conlin_stellarator_2024} to find a solution $\bm{x}_1^*$. In this section we also employ the exponential spectral scaling (ESS) introduced by \cite{jang_exponential_2025} to scale the variables in the problem to respect the expected spectral decay\textcolor{black}{, enabling the optimizations to perform well without the use of a Fourier continuation scheme.} This spectral scaling is employed to scale not the boundary modes as was done in the original work, but instead to scale based on the Fourier-Zernike modes of the equilibrium, as in the constrained optimization approach the degrees of freedom are not just the boundary, but the entire geometry of the equilibrium\footnote{This is in contrast to the typical stage one optimization approach where the ideal MHD force balance constraint is enforced through an equilibrium solve at each step, effectively making the interior of the equilibrium a function of the boundary only and thus the boundary DOFs are the optimization variables in that case.}. The ESS scaling in this work used an exponent of $\alpha=1.2$ and the $L^\infty$ norm. \\
Once $\bm{x}_1^*$ is obtained, the deflated problem is solved, which is the same as the above setup but with an additional deflation constraint:
\begin{equation}
    f_{deflation}:=\prod_iM(\bm{x};\bm{x}_i^*) \leq r
\end{equation}
where $\bm{x}_i^*$ are the prior solutions, and $r$ is some positive constant (chosen to be $1$ here, a scan over larger values revealed no significant difference in results), effectively pushing the optimizer away from some region around each of the already-found solutions. The deflation operator here is applied to the entire equilibrium state (the Fourier-Zernike coefficients for $R_{lmn}$, $Z_{lmn}$ and $\lambda_{lmn}$), with the power and shift parameters $p=2$ and $\sigma=0$. 
We begin from a simple axisymmetric circular torus initial guess with $1$m major radius and $0.125$m minor radius, and proceed with the deflated optimization for 30 iterations. Not every iteration is guaranteed to result in an equilibrium which satisfies all of our constraints after optimization, so we will filter the results to only those which have $\bar{\iota}$ between $0.7$ and $3$, as well as filter for only equilibria with volume averaged force balances below 1\%. 24 equilibria pass these filters, and their boundaries, rotational transform profiles, QS errors, and Boozer surface plots are shown in \autoref{fig:QH-deflated-bdries}, \autoref{fig:QH-deflated-iotas}, \autoref{fig:QH-deflated-QS-errs}, and \autoref{fig:QH-deflated-boozer-plots}, respectively. The results show that the algorithm successfully found a range of equilibria with good QH and varying rotational transform profiles and boundary shapes. \textcolor{black}{\autoref{app:QH-soln-plots} shows additional information on these equilibria, including individual plots of each boundary and different profile quantities.}

However, it can be seen that some similar boundaries and equilibria were found. This is likely due to three reasons: first, the Fourier representation of a single physical boundary shape is not unique (owing to the arbitrariness of the poloidal angle), and thus there may be two state vectors which are "far" from each other in the sense of they have different Fourier coefficients, but the actual physical boundaries they describe are very similar. Use of either an explicit spectral condensation constraint \citep{hirshman_explicit_1998}, or enforcing a poloidal angle representation that is less degenerate \citep{henneberg_representing_2021} may alleviate this issue\footnote{It should be noted that in DESC, spectral condensation is implicitly attained through the freedom afforded to the poloidal stream function, as shown in \cite{panici_desc_2023}, where the spectral width of the \texttt{DESC} solution is found to be comparable to VMEC. This however does not solve the degeneracy issue completely, as multiple different poloidal angles may be similarly spectrally condensed.}. 

\begin{figure}
    \centering
    \includegraphics[width=0.8\linewidth]{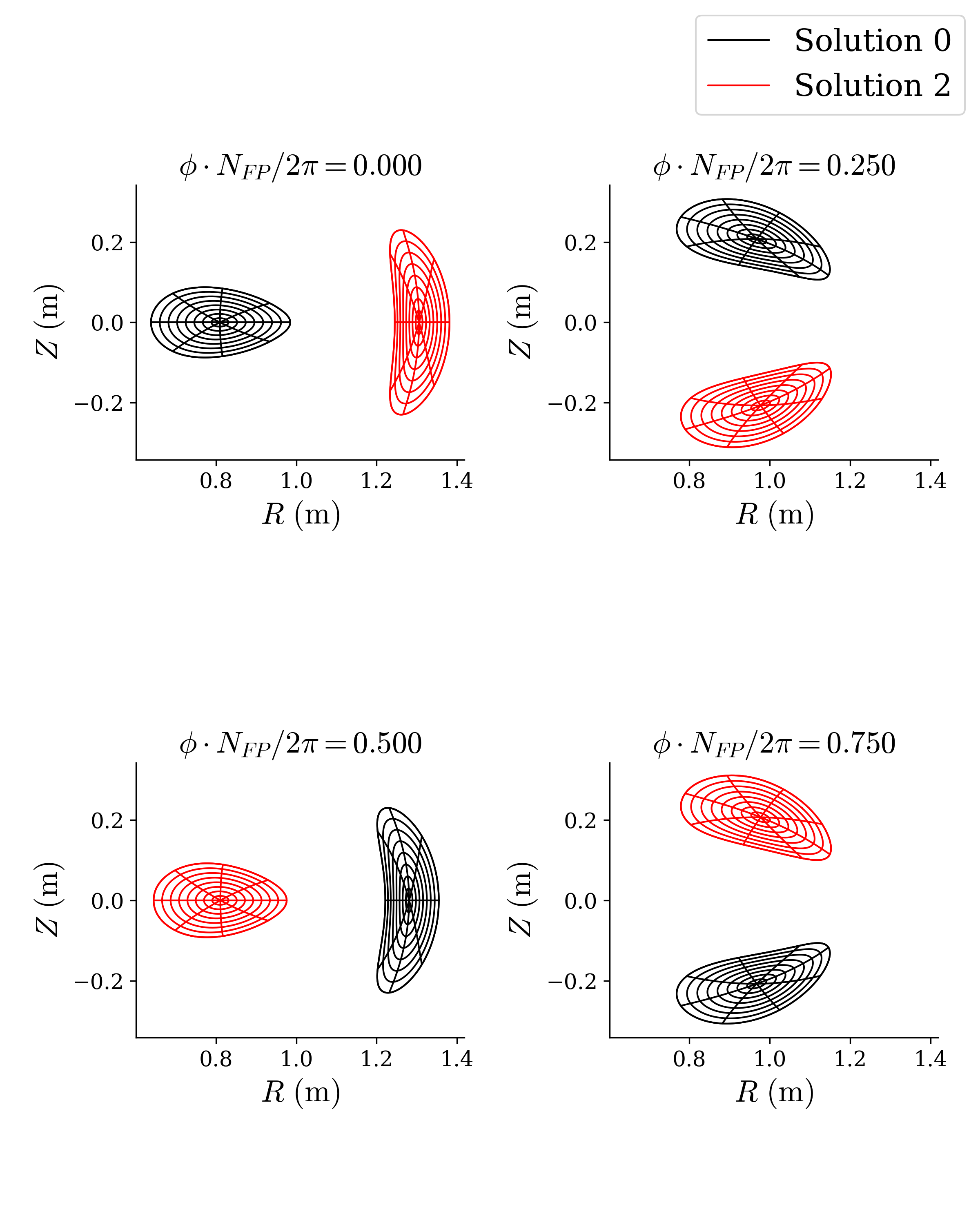}
    \caption{Two of the found QH solutions from \autoref{sec:stage-one-deflation}, the first (index 0, i.e. undeflated) solution and the third (index 2) solution. It can be seen that the equilibria are nearly physically identical asides from a $\pi/N_{FP}$ toroidal rotation, which motivates the inclusion of rotated solutions in the deflation to avoid such degenerate solutions.}    \label{fig:QH-deflated-similar-solutions}
\end{figure}

The second reason is that the same exact physical boundary may, in stellarator-symmetry, be described by the same poloidal angle definition in two different ways, as one may simply shift the toroidal angle in the boundary definition by $\pi/N_{FP}$ to obtain the same physical boundary, but now offset toroidally, similar to the two equilibria in the filtered set shown in \autoref{fig:QH-deflated-similar-solutions}. This problem is not present in \Autoref{sec:nae-multiple-solns}, as fixing the axis and near-axis behavior prevents such shifts. However in \Autoref{sec:helical-core} this problem can and was observed to occur (that is, if one were to run a deflated equilibrium solve using as deflated states the axisymmetric and the helical core state, one would find the helical core state rotated by $\pi$ radians toroidally). A method which could be used to avoid this is to also add the toroidally shifted counterparts of found solutions to the set of deflated states to prevent the other potential parametrization of the same physical solution from being found by the optimizer. Note that if the assumption of stellarator symmetry is relaxed, this problem may be much worse, as now any arbitrary toroidal rotation of the boundary can be found, yielding the same physical solution but potentially being "far" from the deflated solutions in coefficient space. In practice, the relaxation of stellarator symmetry may necessitate more deflation iterations, but this can be ameliorated if one includes one of the toroidally-rotated degenerate solutions whenever one adds a state to the set to deflate against, as shown in Appendix \ref{app:rotations}.

\begin{figure}
    \centering
    \includegraphics[width=1.0\linewidth]{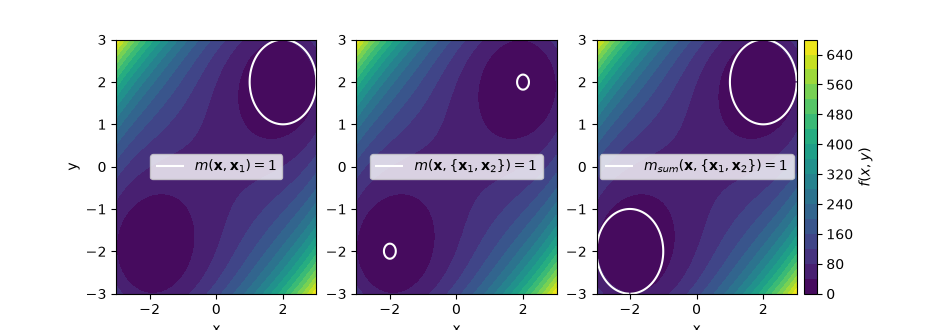}
    \caption{Each panel shows the shaded contours of the function $f(x,y)=\left((x-2)^2+(y-2)^2\right)\left((x+2)^2 + (y+2)^2\right)$, and in each panel in white is the level set corresponding to 1 of (left) the deflation operator acting only on a single deflated minimum $(x,y)=(2,2)$, and the deflation operator acting on two minima, $(2,2)$ and $(-2,-2)$ (middle) with the conventional product-reduction deflation operator, and (right) with the sum-reduction operator. The regions inside the white contours would be infeasible states for the deflated optimization problem, and it can be seen that while the infeasible region around $(2,2)$ is modified when the additional state is added with the product-reduction deflation operator in the middle plot, the sum-reduction operator in the right plot retains the infeasible region of the original deflated state, a more intuitive modification of the optimization landscape.  }\label{fig:sum-vs-prod-deflation}
\end{figure}

The third reason may be due to the form of multiple-deflation used in the problem, along with the choice of $\sigma=0$ for the shift parameter. When deflating multiple solutions, the deflation operators for each individual solution are multiplied together. While this is the method that makes the most sense for nonlinear rootfinding (where one then multiplies the deflation operator $M$ by $\bm{f}$ and seeks $M\bm{f}=0$), in the present case of using the deflation operator as a nonlinear constraint, it may lead to unexpected and unwanted behavior as the number of deflated solution increases. This is because as more solutions are added, the extra cost in the problem due to deflation can be artificially damped by virtue of the existence of more diverse solutions. To put this more concretely,  if we have 2 existing solutions $\bm{x}_1^*$ and $\bm{x}_2^*$, our deflation operator with those two deflated states is given (with $\sigma=0$ and $p=2$) as $m(\bm{x};\{\bm{x}_1^*,\bm{x}_2^*\}) = \frac{1}{|\bm{x}-\bm{x}_1^*|^2}\frac{1}{|\bm{x}-\bm{x}_2^*|^2}$, and our deflation constraint is (as was chosen in this problem) $m\leq r$ and $r=1$. If a given $\bm{x}$ is such that $\frac{1}{|\bm{x}-\bm{x}_1^*|^2}=2$, then that state would be avoided in the first deflation iteration because $m(\bm{x};\bm{x}_1^*)>r$. However, in the second deflation iteration, if that same state $\bm{x}$ now happens to be far from $\bm{x}_2^*$ such that $\frac{1}{|\bm{x}-\bm{x}_2^*|^2}=0.25$, then the deflation cost becomes $m(\bm{x};\{\bm{x}_1^*,\bm{x}_2^*\})=0.5$ which is below the set bound of $r=1$, thus this state is no longer avoided. This is an unintuitive result, and can change the shape of the optimization landscape as the deflation progresses in potentially undesired ways. Not only does the expected and desired change of adding more barriers around other found solutions occur, but now we also are modifying the existing barriers, potentially allowing previously disallowed solutions simply due to the formulation of the constraint. This is schematically shown in \autoref{fig:sum-vs-prod-deflation}, where the white contour is the contour within which the deflation constraint is violated, and it can be seen from the first panel to the second that the addition of the second deflated solution reduces the size of the excluded region around the original deflated solution. One may avoid this unintuitive modification of the landscape by choosing to handle multiple-deflation with a sum instead of a product, i.e.

\begin{equation}\label{eq:sum-reduction}
    m_{sum}(\bm{x};\{\bm{x}_i^*\}) = \sum_i\left(\frac{1}{|\bm{x}-\bm{x}_i^*|^2} + \sigma \right)
\end{equation}

or, to avoid the accretion of the cost due simply to the accumulation of $\sigma$, using only a single shift (as first proposed by \cite{riley_deflation_2025}):
\begin{equation}
    m_{sum}(\bm{x};\{\bm{x}_i^*\}) =  \sigma + \sum_i\frac{1}{|\bm{x}-\bm{x}_i^*|^p} 
\end{equation}

 This form of the multiple deflation operator, which we will call the "sum-reduction" deflation operator, will be used in the next section on stage-two optimization to avoid the problems outlined in this section. The rightmost panel of \autoref{fig:sum-vs-prod-deflation} shows that with the sum-reduction deflation operator, the excluded region no longer shrinks in size when the second deflated solution is included. One could also choose to increase the bound proportional to the amount of solutions being deflated, offering a more controlled way to modify the infeasible region sizes compared to the product-reduction operator.

\subsection{Stage-two Filamentary Coil Optimization}\label{sec:stage-two-deflation}

Like stage-one optimizations, stage-two filamentary coil optimizations are notoriously nonconvex problems that can have many local minima. Once again, we will apply the deflation operator as a constraint during optimization to find multiple local minima, this time using the "sum-reduction" deflation operator of \autoref{eq:sum-reduction} to deal with multiple deflations. We will find coils for the \textcolor{black}{two-field period stellarator-symmetric vacuum "precise QA" stellarator of \cite{landreman_magnetic_2022}, which has an average major radius of approximately 1 meter, an aspect ratio of 6, and an on-axis magnetic field strength of 1 Tesla. We find coils} using a stellarator-symmetric coilset with four coils per half-field period. We parametrize the coil geometry as filamentary curves with Fourier series for the X,Y and Z components of the curves, as first done by FOCUS \citep{zhu_new_2018,zhu_designing_2018}, using Fourier modes up to fifth order in the curve parameter $t$. The least-squares augmented Lagrangian approach in \texttt{DESC} is again used to perform the coil optimization, similar to the approach taken in \cite{gil_augmented_2025}. The ESS scaling is also applied here to the Fourier components of the coil geometry, \textcolor{black}{again motivated by the desire to eliminate the need for Fourier continuation}. The objective being minimized is the quadratic flux on the last closed flux surface:

\begin{equation}
    f_{B_n} = \int_S (\bm{B}\cdot\bm{n})^2dA 
\end{equation}

where $\bm{B}$ is the coil-generated magnetic field, $\bm{n}$ is the unit surface normal, $S$ is the last closed flux surface and $dA$ is the local surface element. This objective is then minimized subject to a set of constraints on the coil geometries and currents, specifically:
\begin{itemize}
    \item The coil currents are constrained such that their sum matches the necessary poloidal linking current demanded by the equilibrium, ensuring the field magnitude matches and also avoiding the trivial solution of zero current \citep{panici_surface_2025}.
    \item The coil-coil minimum distance is bounded to $d_{cc}\geq 0.09$ m.
    \item The coil-plasma minimum distance is bounded to $d_{cp}\geq 0.2$ m.
    \item The pointwise coil curvature $\kappa$ is bounded to be between $-8~\text{m}^{-1}$ and $8~\text{m}^{-1}$ to discourage excessive coil shaping.
    \item Each individual coil length is bounded to be less than 6.5 m. This differs from the conventional sum of lengths constrained to a target length used in stage-two optimization \citep{zhu_new_2018,wechsung_precise_2022,kaptanoglu_reactor-scale_2025, gil_augmented_2025}. Constraining individual coil lengths was found to work better, as when only the length sum is constrained, feasible solutions include coilsets where a single coil gets very long, which is usually not a desired solution engineering-wise.
    \item The variance of the incremental coil arclength, $var(\bm{\Gamma}'(t))$ is constrained to be zero, similar to the constraint proposed by \cite{wechsung_precise_2022}, where it is argued that such a constraint may reduce the inherent degeneracy in the curve parametrization. This is particularly important for this work because, as previously discussed, degeneracy in parametrization may hinder deflation. Practically, this variance will not be zero, but will at least be penalized to remain small.
\end{itemize}

A grid of 101 points evenly spaced in the curve parameter $t$ is used to discretize the coil for purposes of the Biot-Savart law and any other coil geometry computations, and a stellarator-symmetric grid of 51 poloidal and  101 toroidal points on the plasma surface is used for evaluating the quadratic flux and coil-plasma distance objectives. An additional deflation constraint is then added using the sum-reduction form given in \autoref{eq:sum-reduction}, with $\sigma=0$ and $p=2$, and the part of state used for deflation will be the Fourier coefficients of the coil geometries. The value of this deflation metric is constrained to be less than 1:

\begin{equation}
    m_{sum}(\bm{x};\{\bm{x}_i^*\}) \leq 1
\end{equation}

\begin{figure}
    \centering
    \includegraphics[width=0.85\linewidth]{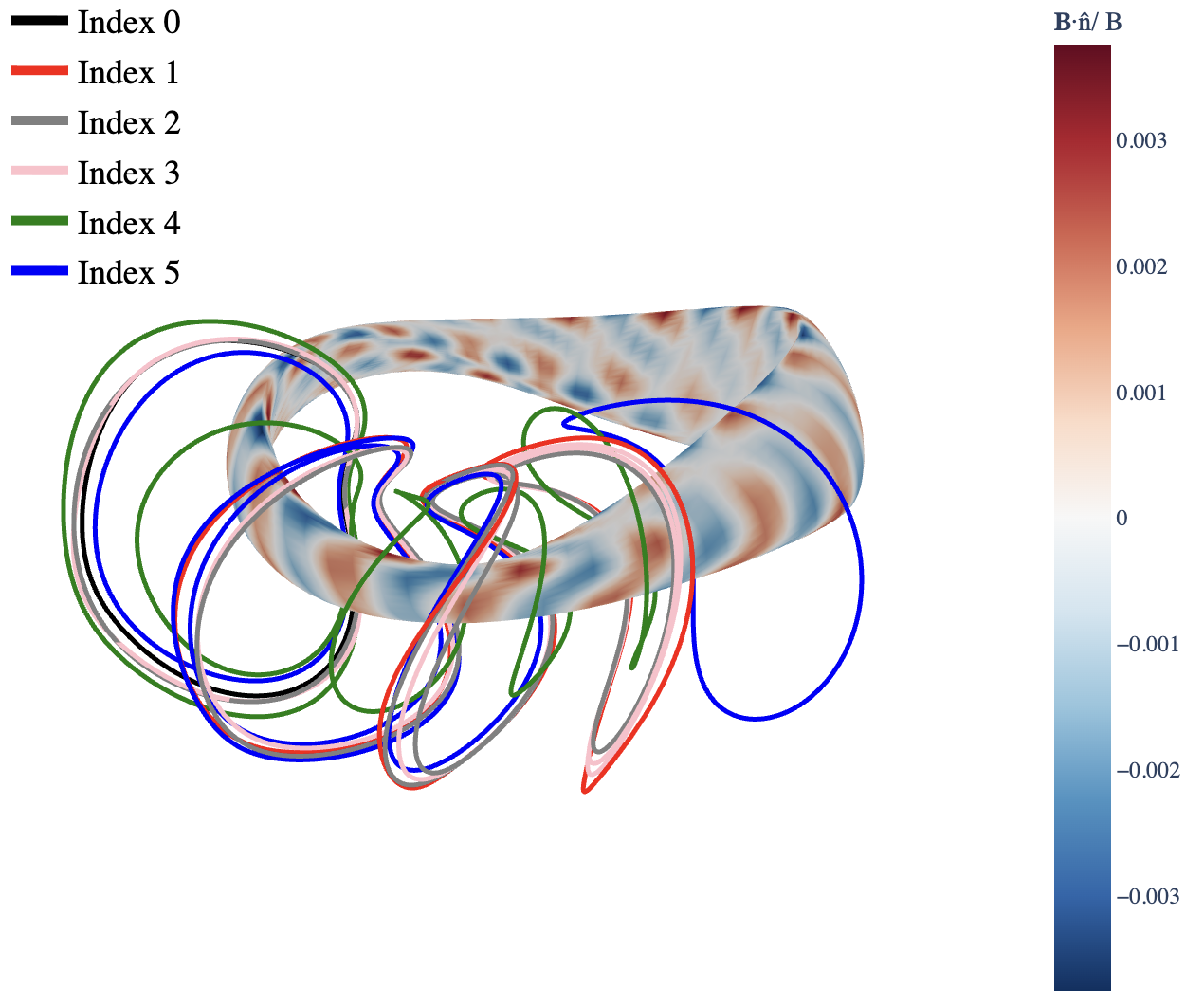}
    \caption{The unique coils in each of the 6 optimized filamentary coilsets for the vacuum precise QA configuration \textcolor{black}{\citep{landreman_magnetic_2022}}. The normal field error from the first optimized coilset is shown on the equilibrium surface.}
    \label{fig:deflate-coils-3d}
\end{figure}

We first run the optimization without the deflation constraint, then  perform five additional optimizations including the deflation constraint. We find six different coilsets, which we show visually in \autoref{fig:deflate-coils-3d} and compare the properties of each in \autoref{tab:deflation-coils}.

It can be seen that all but one of the coilsets satisfy the constraints and achieve good normal field errors \citep{kappel_magnetic_2024}. \textcolor{black}{ While these coilsets may not push the state-of-the-art best coil results for this configuration \citep{wechsung_precise_2022} (though their performance is quite respectable for having only a fifth-order Fourier series parametrization), one of the key takeaways of this optimization is that the second coilset performs the best in terms of normal field error, not the first coilset. This indicates that the deflation method worked well at traversing the landscape, as for the given parameters of the optimization, the optimization for the second coilset (the first optimization that had a nontrivial deflation constraint, which pushed it away from the first coilset) ended up landing in a better local minimum in terms of normal field error (as well as having the highest coil-plasma distance, and comparable curvature to the other coilsets).}
\textcolor{black}{Another interesting point is that the} index-4 coilset with distinctly worse normal field error is the only one which failed to satisfy the constraints, including the deflation constraint, indicating that given the initial guess and optimizer settings used, it could not find a feasible solution. It is especially interesting that physically, this coilset (in green in \autoref{fig:deflate-coils-3d}) seems to be the most visually distinct from the others, showing that the parametrization degeneracy issue is perhaps not entirely solved with the use of the arclength variance penalty included in this problem.

\begin{table}
    \centering
    \begin{tabular}{c c c c c c c}
Coilset & $\frac{\langle B_n \rangle}{|\bm{B}|}$ & $d_{cc}$ [m] & $d_{cp}$ [m] & $\kappa_{max}$ [$\text{m}^{-1}$] & Max Coil Length [m] & $Var(\bm{\Gamma}'(t))$ [$\text{m}^2$] \\ \hline
0 & 1.011e-03 & 0.1465 & 0.2684 & 3.5612 & 4.8419 & 2.948e-03  \\ \hline
1 & 6.719e-04 & 0.1304 & 0.2930 & 3.5944 & 5.2340 & 1.911e-03  \\ \hline
2 & 1.393e-03 & 0.1545 & 0.2752 & 3.3720 & 4.8624 & 4.319e-03  \\ \hline
3 & 1.263e-03 & 0.1495 & 0.2731 & 3.4583 & 4.8911 & 4.013e-03  \\ \hline
4 & 2.327e-02 & 0.1443 & 0.2001 & 5.0100 & 3.6251 & 1.702e-02  \\ \hline
5 & 1.806e-03 & 0.1541 & 0.2525 & 3.3786 & 4.5227 & 5.518e-03  \\ \hline
    \end{tabular}
    \caption{Figures of merit for the 6 optimized coilsets found through the deflation algorithm, where $\frac{\langle B_n \rangle}{|\bm{B}|}$ denotes the average normalized normal field error over the plasma surface, with the normalizing field as the coil-generated magnetic field. }
    \label{tab:deflation-coils}
\end{table}






\section{Conclusions and Future Work}
\label{sec:conclusions}
In this work, the deflation method was introduced and successfully applied to stellarator equilibrium and optimization problems to find multiple solutions and local minima. The deflation method offers an additional tool for exploring the complex objective landscapes which characterize both stellarator optimization and non-axisymmetric equilibrium solves, beyond the conventional approaches of launching the same optimization with many different initial guesses (which can be wasteful, as many may converge to the same or similar local minima) or tinkering with objective weights (which can be infeasible to scan systematically when dealing with many objectives). The ease of use for this tool must also be emphasized: in each application shown, the deflation algorithm simply was a repeated call to the original optimization or equilibrium problem with a modified objective or augmented with an additional constraint, requiring minimal extra effort to implement and yielding excellent results.\par
For equilibrium studies, this work showed that deflation methods can be leveraged to find multiple local minima to the near-axis-constrained equilibrium problem, revealing entire families of possible initial guesses which share desirable core characteristics. Deflated equilibrium solves were also used to obtain perturbed helical core equilibria without the need to presciently prescribe the perturbation, showing deflation's utility in finding bifurcated equilibrium states. These techniques can be applied to any equilibrium problem, so one could imagine modifying another equilibrium code like the Stepped Pressure Equilibrium Code \citep{hudson_computation_2012,qu_stepped_2020} to include such methods, where the different equilibrium model and lack of nested surface assumption may permit more interesting bifurcated states to be found.\par
In stellarator optimization, deflation was used to find multiple distinct and attractive local minima to the stage-one QH optimization problem and the stage-two coil optimization problem, utilizing both conventional deflation operators as well as a novel sum-reduction deflation operator. This showcases deflation's potential to aid in searching the complex, multi-objective landscapes which characterize stellarator optimization. The deflated optimization in this work is straightforward to implement, and can be easily incorporated into other optimization software \citep{landreman_simsopt_2021,zhu_new_2018,drevlak_optimisation_2019}. Additionally, given that stage-one and stage-two optimization problems are nonconvex, it is only natural that the combined single-stage problem shares this characteristic. The sensitivity of the result to the weights of the objectives in single-stage optimization is even higher \citep{jorge_single-stage_2023,jorge_simplified_2024}, and deflation may be useful to find multiple local minima in this case. \par
There are multiple avenues of improvement to the deflation methods shown in this work. While the specific implementation chosen for this paper is effective and yields positive results, an improved formulation for least-squares minimization problems following that of \cite{riley_deflation_2025} would likely yield a more robust deflation method that can successfully converge to higher numbers of distinct solutions, and so will be pursued in future work. These improvements would come from the modified deflated least-squares algorithm being allowed to choose when to take steps which reduce the deflated problem residual and when to take steps which reduce only the undeflated residual when far from known solutions, something which the current implementation's baking the deflation into the objective or constraints does not allow.\par 
Combating the non-uniqueness of the Fourier representation of both coils and surfaces would also improve deflation. While extremely versatile, the Fourier representation suffers from both the issue of parametrization degeneracy, as highlighted earlier in this work, and allowing non-physical, self-intersecting surfaces, both of which render the deflated optimization problem more difficult. The stellarator optimization problem itself would benefit from advances in surface parametrization which improve upon these flaws, and these benefits would apply as well to the effectiveness of deflation in ensuring that mathematically distinct deflated solutions found are also physically distinct. Recent work by \cite{jang_exponential_2025} on incorporating the expected exponential scaling of such a representation into the optimization process already has been shown to help with these aspects, and indeed was utilized in this paper, but further work is still warranted. For example, the exponential decay of the Fourier representation means that the unweighted $L^2$ norm used in this deflation work is dominated by the lower-order modes of the state, essentially discouraging solutions which may have similar low order shaping but significantly different higher order shaping. One could instead use an ESS-weighted norm which would more equally weigh the higher-order parts of the state, allowing such solutions to be found.\par
It could also be interesting to explore deflation not on the state directly, but rather some reduced state or some derived quantities of the state, which may avoid the potential degeneracies associated with the Fourier representation. \textcolor{black}{This was already explored preliminarily in \Autoref{sec:helical-core}, where the deflation was performed on the magnetic axis.} As a further example, one could imagine using an equilibrium's mean rotational transform, minimum magnetic well, and minimum $L_{\nabla B}$ on the surface as a sort of reduced state for a stellarator. While certainly not a unique identifier for a stellarator, if one is considering tradeoffs in an optimization space, distinct solutions of interest would certainly have differing values for these quantities. Similarly, one could take inspiration from NAE theory and use just an equilibrium's axis shape and other near-axis quantities, which from the NAE perspective have been shown to discriminate between distinct designs \citep{rodriguez_constructing_2023,landreman_mapping_2022}. \textcolor{black}{Other methods of exploring the NAE-constrained equilibrium landscape could also be pursued, such as taking parametric combinations of the solutions found in this work and attempting to solve for equilibrium at intermediate points between equilibria, or using principal-component-analysis to find a reduced basis within which to search for solutions.}\par
Finally, further scans over the deflation method hyperparameters and operator types would be useful to probe the sensitivity of the method's effectiveness to the chosen hyperparameters. The hyperparameters $p$, $\sigma$ (for deflated equilibrium problem), and $r$ (the additional hyperparameter for  deflated constrained optimization) can be tuned to modify the space of found solutions, as well as the deflation operator types: sum versus product, single-shift versus multiple-shift, and norm-deflation versus exponential-deflation \citep{riley_deflation_2025}. If there is significant sensitivity, as was suggested by preliminary scans during this work, it would actually be a positive outcome, as one may merely use different hyperparameter choices to more broadly search the optimization or solution landscape. Deflation could also be paired with conventional methods of multiple initial conditions or systematic weight scans to increase the effectiveness of each.

\section{Data Availability}
The deflation methods shown in this paper are available at the \texttt{DESC} Github repository \citep{DESC}. The scripts and figures used are freely available on Princeton Data Commons with DOI \href{https://doi.org/10.34770/swdq-2f72}{https://doi.org/10.34770/swdq-2f72} \citep{panici_dario_2026_dataset_final}.

\section{Acknowledgments}
One of the authors (D.P.) acknowledges useful discussions with Elizabeth Paul, Rogerio Jorge, Alan Kaptanoglu, Frank Fu, and Pedro Gil on possible applications of deflation techniques to stellarator optimization. 
This work is funded through the SciDAC program by the US Department of Energy, Office of Fusion Energy Science and Office of Advanced Scientific Computing Research under contract number DE-AC02-09CH11466, DE-SC0022005, and by the Simons Foundation/SFARI (560651). The United States Government retains a non-exclusive, paid-up, irrevocable, world-wide license to publish or reproduce the published form of this manuscript, or allow others to do so, for United States Government purposes.

\section{Declaration of Interest}
The authors have no conflicts of interest to declare, and no AI tools were used in this work.

\bibliographystyle{jpp}
\bibliography{main, bib}

\appendix

\section{Further Physics Metrics for the Multiple QA NAE-constrained Solutions}\label{app:nae-soln-plots}
In this appendix we share some additional plots and values of various metrics for the 25 QA first-order NAE-constrained solutions shown in \ref{sec:nae-multiple-solns}. \textcolor{black}{Table \ref{tab:qa-info} has tabulated information for each equilibrium, including the on-axis and edge iota, the volume-averaged normalized force balance error, the max Boozer quasisymmetry error (occuring at the edge for all cases), as well as the mean magnetic well \citep{greene_brief_1998, landreman_magnetic_2020, mercier_equilibrium_1964} and minimum $L_{\nabla B}$ \citep{landreman_figures_2021, landreman_mapping_2022} on the boundary. Each solution found has well-below 1\% average normalized force balance error across the entire volume, indicating that these are each excellent solutions to the ideal MHD equilibrium equations. The magnetic well is mostly negative for the equilibria, indicating instability, however a few solutions have marginal or even stable magnetic wells for part of the volume, which may be good initial points from which to perform further optimizations. Since these solutions were constrained only to first-order to match the NAE, there was no constraint on the second-order quantities of magnetic well and $L_{\nabla B}$, so the values are essentially uncontrolled during the optimization process. Further work is warranted to perform similar optimizations using higher-order NAE-constrained equilibria, which may restrict the space of possible solutions to a subspace with more favorable properties.} \par
\begin{figure}
    \centering
    \includegraphics[width=0.95\linewidth]{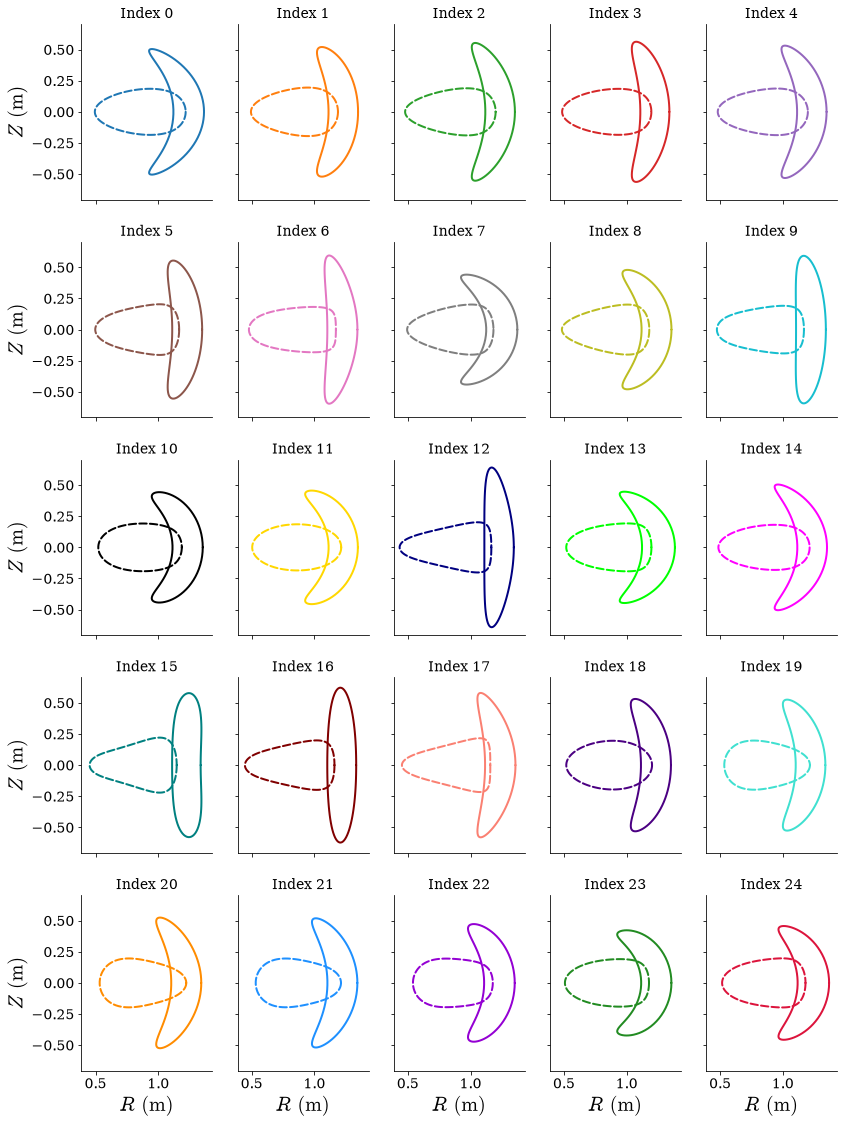}
    \caption{Boundaries at $\phi=0$ (solid) and $\phi=\pi/N_{FP}$ (dashed) for the 25 deflated first-order NAE-constrained equilibria.}
    \label{fig:QA-solns-bdries-grid}
\end{figure}
\begin{figure}
    \centering
    \includegraphics[width=0.95\linewidth]{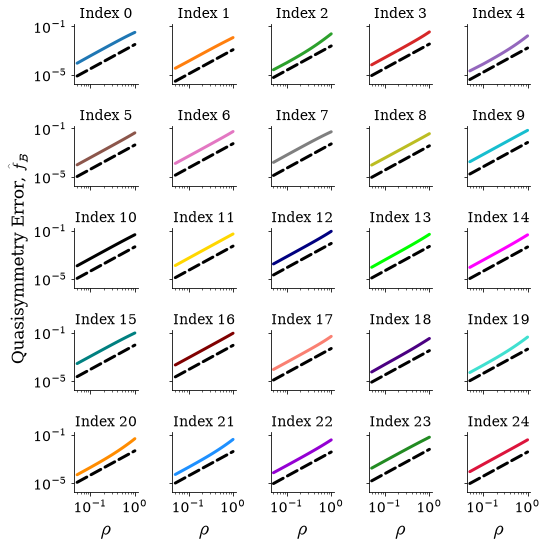}
    \caption{QA errors (as measured by $\hat{f}_B$) versus flux surface label $\rho$ for the 25 deflated first-order NAE-constrained equilibria. Plotted in black is also the expected $\rho^2$ scaling of the QA error for the first-order NAE.}
    \label{fig:QA-solns-QS-errs-grid}
\end{figure}
\begin{figure}
    \centering
    \includegraphics[width=0.95\linewidth]{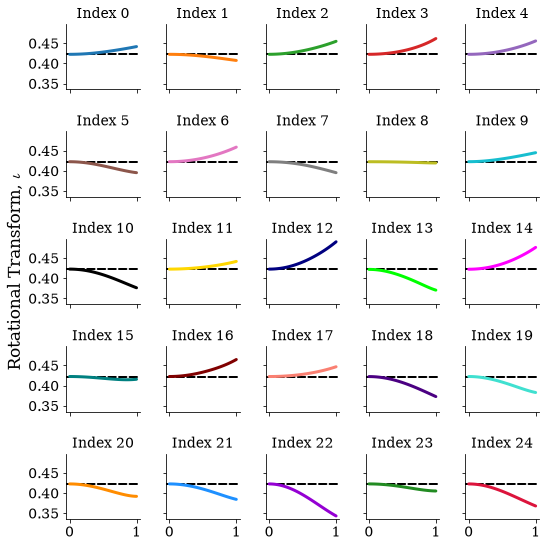}
    \caption{Rotational transform versus flux surface label $\rho$ for the 25 deflated first-order NAE-constrained equilibria. Plotted in black is also the on-axis rotational transform value from the first-order NAE solution.}
    \label{fig:QA-solns-iotas-grid}
\end{figure}
\begin{figure}
    \centering
    \includegraphics[width=0.95\linewidth]{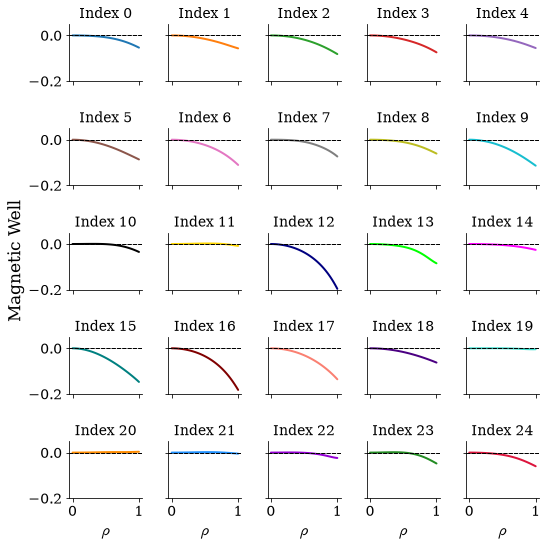}
    \caption{Magnetic well versus flux surface label $\rho$ for the 25 deflated first-order NAE-constrained equilibria. The black dashed line denotes marginal stability, where above the line is associated with a stable magnetic well and below with an unstable magnetic well (magnetic hill).}
    \label{fig:QA-solns-wells-grid}
\end{figure}

\textcolor{black}{Additionally, the boundary shapes at $\phi=0$ and $\phi=\pi/N_{FP}$, QA error profiles, iota profiles, and magnetic well profiles are also shown individually for each equilibrium in Figures \ref{fig:QA-solns-bdries-grid}-\ref{fig:QA-solns-wells-grid}.}



\begingroup
\renewcommand{\arraystretch}{0.7} 
\begin{table}
    \centering
    \begin{tabular}{c c c c c c c}
Eq Index & On-axis $\iota$ & Edge $\iota$ & Normalized $\langle|F|\rangle_{\text{vol}}$ & Mean Magnetic Well & Max $\hat{f}_B$ & Min $L_{\nabla B}$ (m) \\ \hline
0 & 0.42 & 0.44 & $7.89 \times 10^{-5}$ & $-1.45 \times 10^{-2}$ & $3.23 \times 10^{-2}$ & 0.47 \\ \hline
1 & 0.42 & 0.41 & $3.76 \times 10^{-5}$ & $-2.00 \times 10^{-2}$ & $1.20 \times 10^{-2}$ & 0.48 \\ \hline
2 & 0.42 & 0.45 & $4.66 \times 10^{-5}$ & $-2.62 \times 10^{-2}$ & $2.44 \times 10^{-2}$ & 0.49 \\ \hline
3 & 0.42 & 0.46 & $5.72 \times 10^{-5}$ & $-2.36 \times 10^{-2}$ & $3.53 \times 10^{-2}$ & 0.48 \\ \hline
4 & 0.42 & 0.46 & $5.47 \times 10^{-5}$ & $-1.78 \times 10^{-2}$ & $1.66 \times 10^{-2}$ & 0.47 \\ \hline
5 & 0.42 & 0.40 & $6.58 \times 10^{-5}$ & $-3.08 \times 10^{-2}$ & $4.37 \times 10^{-2}$ & 0.48 \\ \hline
6 & 0.42 & 0.46 & $6.28 \times 10^{-5}$ & $-3.37 \times 10^{-2}$ & $5.61 \times 10^{-2}$ & 0.48 \\ \hline
7 & 0.42 & 0.40 & $1.18 \times 10^{-4}$ & $-1.84 \times 10^{-2}$ & $5.35 \times 10^{-2}$ & 0.44 \\ \hline
8 & 0.42 & 0.42 & $1.08 \times 10^{-4}$ & $-1.74 \times 10^{-2}$ & $3.73 \times 10^{-2}$ & 0.46 \\ \hline
9 & 0.42 & 0.45 & $6.40 \times 10^{-5}$ & $-3.89 \times 10^{-2}$ & $6.99 \times 10^{-2}$ & 0.49 \\ \hline
10 & 0.42 & 0.38 & $4.72 \times 10^{-5}$ & $-6.26 \times 10^{-3}$ & $4.54 \times 10^{-2}$ & 0.42 \\ \hline
11 & 0.42 & 0.44 & $4.44 \times 10^{-5}$ & $7.93 \times 10^{-5}$ & $5.23 \times 10^{-2}$ & 0.43 \\ \hline
12 & 0.42 & 0.49 & $1.05 \times 10^{-4}$ & $-5.64 \times 10^{-2}$ & $8.66 \times 10^{-2}$ & 0.50 \\ \hline
13 & 0.42 & 0.37 & $2.34 \times 10^{-4}$ & $-2.31 \times 10^{-2}$ & $4.90 \times 10^{-2}$ & 0.46 \\ \hline
14 & 0.42 & 0.48 & $1.25 \times 10^{-4}$ & $-7.22 \times 10^{-3}$ & $4.43 \times 10^{-2}$ & 0.46 \\ \hline
15 & 0.42 & 0.42 & $6.61 \times 10^{-5}$ & $-5.25 \times 10^{-2}$ & $9.35 \times 10^{-2}$ & 0.48 \\ \hline
16 & 0.42 & 0.46 & $8.71 \times 10^{-5}$ & $-5.31 \times 10^{-2}$ & $8.93 \times 10^{-2}$ & 0.49 \\ \hline
17 & 0.42 & 0.45 & $7.90 \times 10^{-5}$ & $-4.29 \times 10^{-2}$ & $4.91 \times 10^{-2}$ & 0.49 \\ \hline
18 & 0.42 & 0.37 & $7.52 \times 10^{-5}$ & $-2.14 \times 10^{-2}$ & $3.29 \times 10^{-2}$ & 0.49 \\ \hline
19 & 0.42 & 0.38 & $6.39 \times 10^{-5}$ & $-1.01 \times 10^{-3}$ & $4.32 \times 10^{-2}$ & 0.47 \\ \hline
20 & 0.42 & 0.39 & $1.00 \times 10^{-4}$ & $1.56 \times 10^{-3}$ & $4.71 \times 10^{-2}$ & 0.41 \\ \hline
21 & 0.42 & 0.38 & $7.06 \times 10^{-5}$ & $3.17 \times 10^{-4}$ & $4.10 \times 10^{-2}$ & 0.47 \\ \hline
22 & 0.42 & 0.34 & $5.49 \times 10^{-5}$ & $-5.11 \times 10^{-3}$ & $3.74 \times 10^{-2}$ & 0.46 \\ \hline
23 & 0.42 & 0.41 & $7.70 \times 10^{-5}$ & $-8.22 \times 10^{-3}$ & $6.22 \times 10^{-2}$ & 0.40 \\ \hline
24 & 0.42 & 0.37 & $7.72 \times 10^{-5}$ & $-1.67 \times 10^{-2}$ & $3.77 \times 10^{-2}$ & 0.46 \\ \hline
    \end{tabular}
    \caption{Quantities of interest for all 25 NAE-constrained equilibria solved in \autoref{sec:nae-multiple-solns}.}
    \label{tab:qa-info}
\end{table}
\endgroup

    


\section{Further Physics Metrics for the Multiple QH Optimized Solutions}\label{app:QH-soln-plots}
\textcolor{black}{In this appendix we share some additional plots and values of various metrics for the 24 QH stage-one-optimized solutions shown in \ref{sec:stage-one-deflation}. Table \ref{tab:qh-info} has tabulated information for each equilibrium, including the on-axis and edge iota, the volume-averaged normalized force balance error, the max Boozer quasisymmetry error (occuring at the edge for all cases), as well as the mean magnetic well \citep{greene_brief_1998, landreman_magnetic_2020, mercier_equilibrium_1964} and minimum $L_{\nabla B}$ \citep{landreman_figures_2021, landreman_mapping_2022} on the boundary. Each solution found has well-below 1\% average normalized force balance error across the entire volume, indicating that these are each excellent solutions to the ideal MHD equilibrium equations. The magnetic well is negative for all optimized equilibria, indicating instability, however stability was not included as part of the optimization objective.} \par
\begin{figure}
    \centering
    \includegraphics[width=0.95\linewidth]{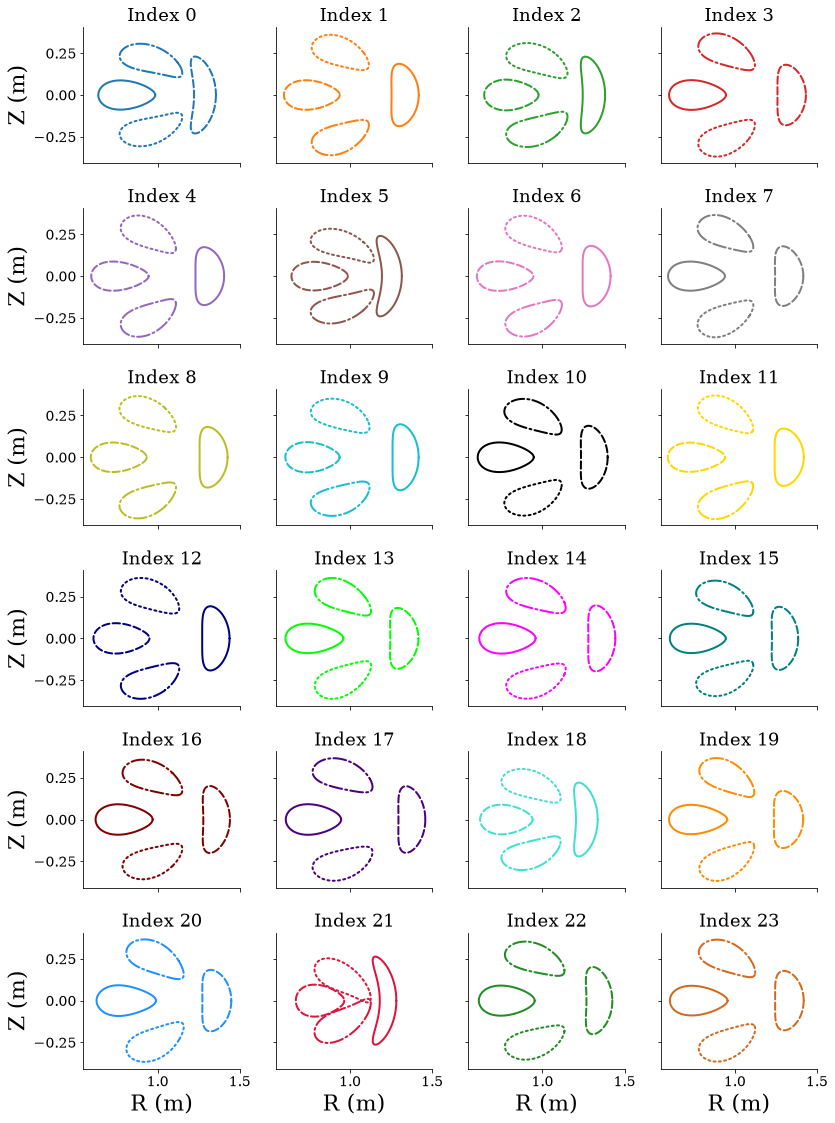}
    \caption{Boundaries at $\phi=0$ (solid), $\phi=\pi/2N_{FP}$ (dash-dotted),  $\phi=\pi/N_{FP}$ (dashed) and  $\phi=3\pi/2N_{FP}$ (dotted) for the 24 QH equilibria optimized in \autoref{sec:stage-one-deflation}.}
    \label{fig:QH-solns-bdries-grid}
\end{figure}
\begin{figure}
    \centering
    \includegraphics[width=0.95\linewidth]{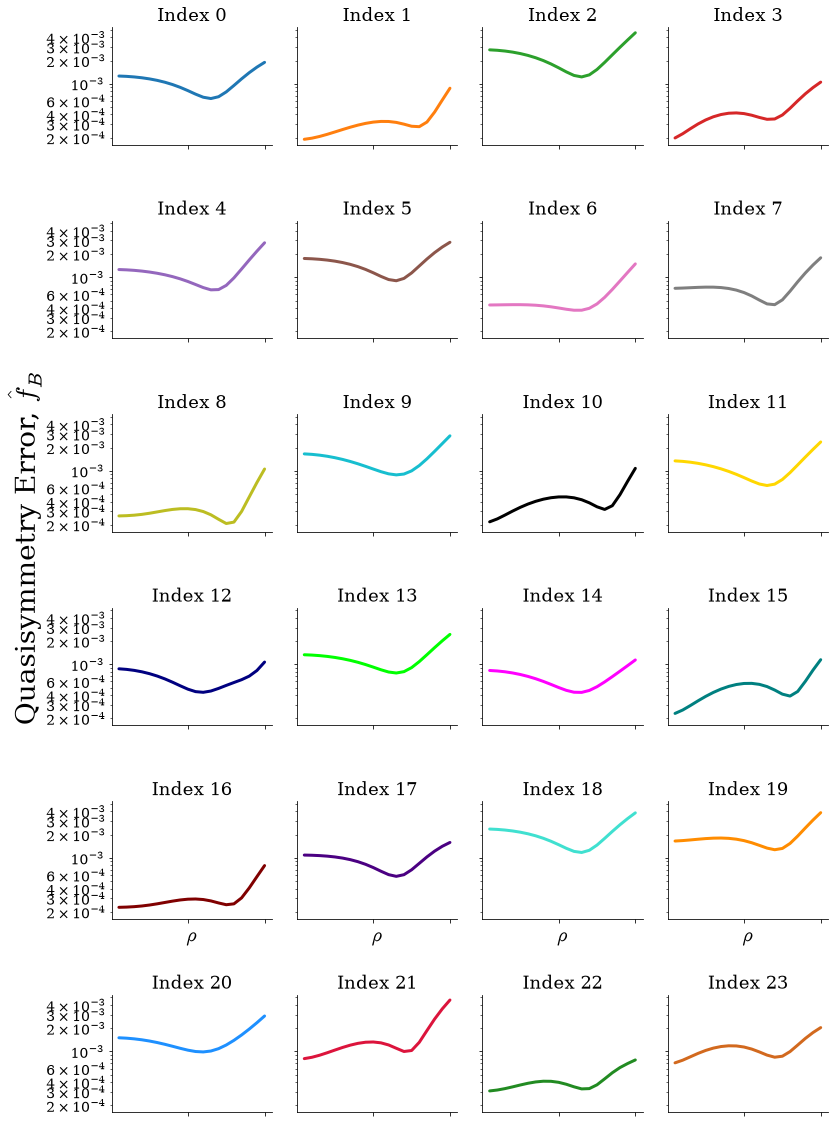}
    \caption{QH errors (as measured by $\hat{f}_B$) versus flux surface label $\rho$ for the 24 QH equilibria optimized in \autoref{sec:stage-one-deflation}. }
    \label{fig:QH-solns-QS-errs-grid}
\end{figure}
\begin{figure}
    \centering
    \includegraphics[width=0.95\linewidth]{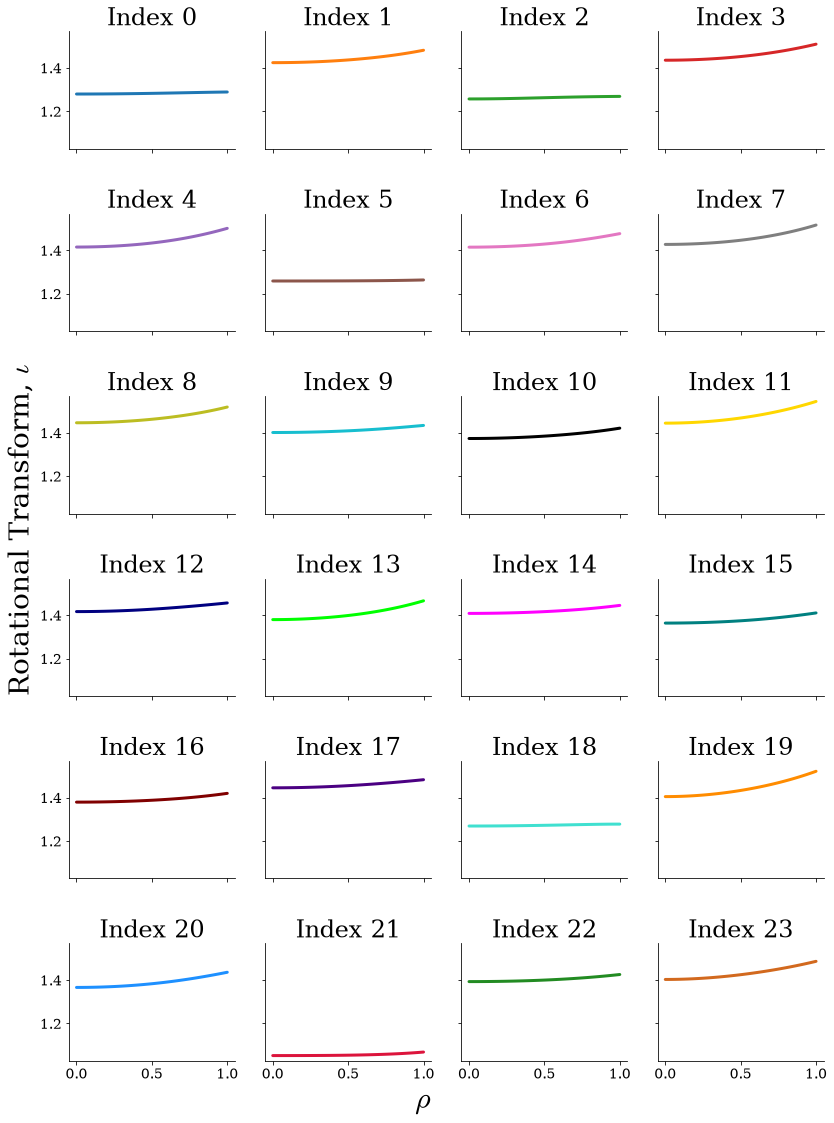}
    \caption{Rotational transform versus flux surface label $\rho$ for the 24 QH equilibria optimized in \autoref{sec:stage-one-deflation}.}
    \label{fig:QH-solns-iotas-grid}
\end{figure}

\textcolor{black}{Additionally, the boundary shapes at $\phi=0, \pi/(2N_{FP}), \pi/(N_{FP}), \text{ and } 3\pi/(2N_{FP})$, QH error profiles, and iota profiles, are also shown individually for each equilibrium in Figures \ref{fig:QH-solns-bdries-grid}-\ref{fig:QH-solns-iotas-grid}.}



\begingroup
\renewcommand{\arraystretch}{0.7} 
\begin{table}
    \centering
    \begin{tabular}{c c c c c c c}
Eq Index & On-axis $\iota$ & Edge $\iota$ & Normalized $\langle|F|\rangle_{\text{vol}}$ & Mean Magnetic Well & Max $\hat{f}_B$ & Min $L_{\nabla B}$ (m) \\ \hline
0 & 1.28 & 1.29 & $5.38 \times 10^{-4}$ & $-8.35 \times 10^{-3}$ & $1.90 \times 10^{-3}$ & 0.29 \\ \hline
1 & 1.42 & 1.48 & $3.50 \times 10^{-4}$ & $-1.96 \times 10^{-2}$ & $8.76 \times 10^{-4}$ & 0.33 \\ \hline
2 & 1.26 & 1.27 & $1.32 \times 10^{-3}$ & $-1.00 \times 10^{-2}$ & $4.58 \times 10^{-3}$ & 0.30 \\ \hline
3 & 1.44 & 1.51 & $5.26 \times 10^{-4}$ & $-2.25 \times 10^{-2}$ & $1.05 \times 10^{-3}$ & 0.34 \\ \hline
4 & 1.42 & 1.50 & $6.54 \times 10^{-4}$ & $-2.34 \times 10^{-2}$ & $2.80 \times 10^{-3}$ & 0.34 \\ \hline
5 & 1.26 & 1.26 & $5.06 \times 10^{-4}$ & $-6.73 \times 10^{-3}$ & $2.84 \times 10^{-3}$ & 0.27 \\ \hline
6 & 1.41 & 1.48 & $6.20 \times 10^{-4}$ & $-2.09 \times 10^{-2}$ & $1.49 \times 10^{-3}$ & 0.34 \\ \hline
7 & 1.43 & 1.52 & $3.44 \times 10^{-4}$ & $-2.36 \times 10^{-2}$ & $1.79 \times 10^{-3}$ & 0.33 \\ \hline
8 & 1.45 & 1.52 & $3.26 \times 10^{-4}$ & $-2.21 \times 10^{-2}$ & $1.06 \times 10^{-3}$ & 0.34 \\ \hline
9 & 1.40 & 1.43 & $6.81 \times 10^{-4}$ & $-1.60 \times 10^{-2}$ & $2.84 \times 10^{-3}$ & 0.33 \\ \hline
10 & 1.37 & 1.42 & $4.65 \times 10^{-4}$ & $-1.71 \times 10^{-2}$ & $1.08 \times 10^{-3}$ & 0.33 \\ \hline
11 & 1.44 & 1.54 & $5.61 \times 10^{-4}$ & $-2.56 \times 10^{-2}$ & $2.37 \times 10^{-3}$ & 0.34 \\ \hline
12 & 1.42 & 1.46 & $7.17 \times 10^{-4}$ & $-1.82 \times 10^{-2}$ & $1.07 \times 10^{-3}$ & 0.34 \\ \hline
13 & 1.38 & 1.47 & $4.50 \times 10^{-4}$ & $-2.16 \times 10^{-2}$ & $2.45 \times 10^{-3}$ & 0.33 \\ \hline
14 & 1.41 & 1.45 & $4.28 \times 10^{-4}$ & $-1.68 \times 10^{-2}$ & $1.14 \times 10^{-3}$ & 0.34 \\ \hline
15 & 1.36 & 1.41 & $5.29 \times 10^{-4}$ & $-1.64 \times 10^{-2}$ & $1.15 \times 10^{-3}$ & 0.33 \\ \hline
16 & 1.38 & 1.42 & $3.91 \times 10^{-4}$ & $-1.60 \times 10^{-2}$ & $7.95 \times 10^{-4}$ & 0.33 \\ \hline
17 & 1.45 & 1.48 & $4.19 \times 10^{-4}$ & $-1.71 \times 10^{-2}$ & $1.58 \times 10^{-3}$ & 0.34 \\ \hline
18 & 1.27 & 1.28 & $9.10 \times 10^{-4}$ & $-1.01 \times 10^{-2}$ & $3.83 \times 10^{-3}$ & 0.29 \\ \hline
19 & 1.40 & 1.52 & $6.57 \times 10^{-4}$ & $-2.51 \times 10^{-2}$ & $3.86 \times 10^{-3}$ & 0.34 \\ \hline
20 & 1.37 & 1.44 & $9.71 \times 10^{-4}$ & $-2.09 \times 10^{-2}$ & $2.88 \times 10^{-3}$ & 0.05 \\ \hline
21 & 1.05 & 1.07 & $2.82 \times 10^{-3}$ & $-4.51 \times 10^{-3}$ & $4.64 \times 10^{-3}$ & 0.24 \\ \hline
22 & 1.39 & 1.43 & $4.46 \times 10^{-4}$ & $-1.53 \times 10^{-2}$ & $7.75 \times 10^{-4}$ & 0.33 \\ \hline
23 & 1.40 & 1.49 & $7.99 \times 10^{-4}$ & $-2.16 \times 10^{-2}$ & $2.04 \times 10^{-3}$ & 0.34 \\ \hline
    \end{tabular}
    \caption{Quantities of interest for all 24 QH equilibria optimized in \autoref{sec:stage-one-deflation}.}
    \label{tab:qh-info}
\end{table}
\endgroup

\section{Deflation Cost for Degenerate Toroidal Rotations Without Stellarator Symmetry}\label{app:rotations}

\begin{figure}
    \centering
    \includegraphics[width=0.95\linewidth]{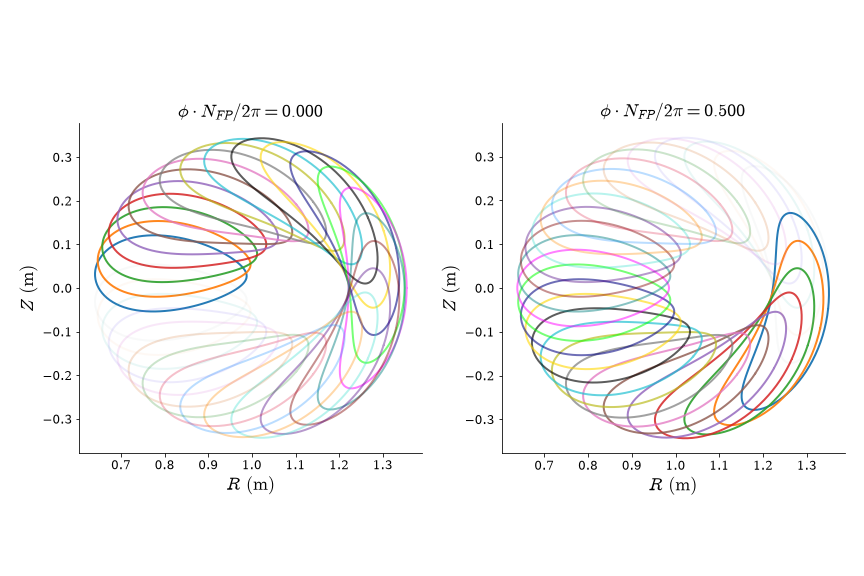}
    \caption{The boundaries of a QH solution from \autoref{sec:stage-one-deflation} along with a series of toroidal rotation of the same solution, shown at two different toroidal angles. Note that all but the rotation which corresponds to $\Delta \zeta_i = \pi/N_{FP}$ are non-stellarator symmetric and thus are not possible to attain when stellarator symmetry is enforced.}
    \label{fig:QH-rotations}
\end{figure}

\begin{figure}
    \centering
    \includegraphics[width=0.95\linewidth]{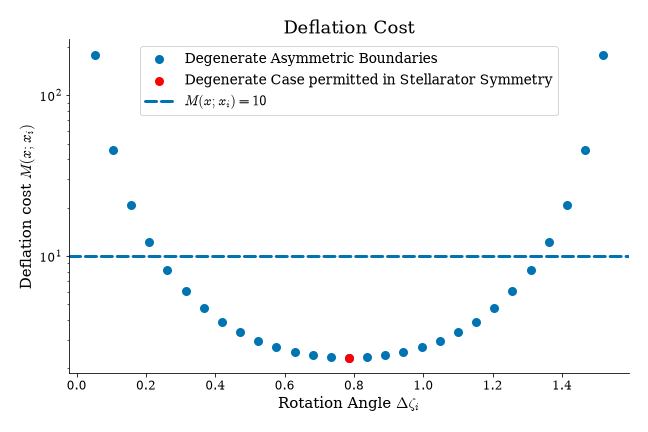}
    \caption{The deflation metric cost for the rotated QH rotated boundaries when deflating against the original solution's boundary. Note that the solutions nearby $\Delta \zeta_i=0$ and $\Delta \zeta_i=2\pi/N_{FP}$ are still sufficiently avoided (i.e. their cost is above the chosen threshold of $M(x;x_i)=10$). The stellarator-symmetric rotation's cost  is marked in red.}
    \label{fig:QH-rotations-cost}
\end{figure}

\begin{figure}
    \centering
    \includegraphics[width=0.75\linewidth]{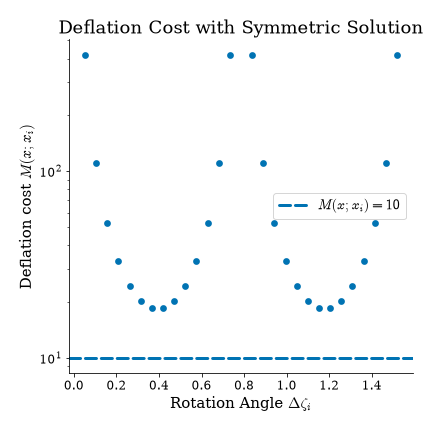}
    \caption{The deflation metric cost for the rotated QH rotated boundaries when deflating against both the original solution's boundary and its $\pi/N_{FP}$ toroidal rotation. Note that now all possible rotations are sufficiently avoided, despite only two of the solutions being included in the deflation.}
    \label{fig:QH-rotations-cost-with-sym}
\end{figure}

In this appendix a short investigation of the landscape of the deflation metric cost over degenerate toroidal rotations is investigated. One of the QH solutions from \ref{sec:stage-one-deflation} is taken and then rotated around toroidally by $\Delta \zeta_i=2\pi i/(nN_{FP})$ for $i\in[1,2,..,n)$ and $n=30$, the cross-sections of which at two toroidal angles are shown in \autoref{fig:QH-rotations}. The deflation metric $M(x;x_i)$ cost of each toroidally rotated state when deflating against the original solution's boundary Fourier coefficients (with $p=2$ and $\sigma=0$) is shown in \autoref{fig:QH-rotations-cost}, along with a horizontal line at a chosen threshold cost of $M=10$. In this appendix we consider solutions with cost $M>10$ to be sufficiently avoided by the deflation operator. It can be seen that in this case, the deflation metric does avoid some of the solutions nearby to the original deflated solution, illustrating the sort of peak in the cost landscape which is inserted at the point of the deflated state. However, toroidal rotations further away in parameter space near $\pi/N_{FP}$ have lower cost and may not be avoided with this deflation metric. To alleviate this, one can choose to always add both the desired state to avoid as well as it $\pi/N_{FP}$ toroidal rotation, which will place two barriers against (toroidally) degenerate solutions, resulting in an avoidance of each possible toroidal degeneracy as shown in \autoref{fig:QH-rotations-cost-with-sym}. 

\end{document}